\documentclass[fleqn,usenatbib]{mnras}

\usepackage[T1]{fontenc}

\DeclareRobustCommand{\VAN}[3]{#2}
\let\VANthebibliography\thebibliography
\def\thebibliography{\DeclareRobustCommand{\VAN}[3]{##3}\VANthebibliography}

\usepackage{graphicx}	
\usepackage{amsmath}	
\usepackage{amssymb}	
\usepackage{dblfloatfix}
\usepackage{caption}
\usepackage{nccmath}
\usepackage{comment}
\usepackage{multirow}
\usepackage{newtxtext,newtxmath}

\usepackage[justification=centering]{caption}

\title[]{Clouds accreting from the IGM are not able to feed the star formation of low-redshift disc galaxies}

\author[A. Afruni et al.]{
Andrea Afruni,$^{1,2}$\thanks{E-mail: andreaafruni@gmail.com}
Gabriele Pezzulli,$^{2}$
Filippo Fraternali,$^{2}$\ \&
Asger Gr{\o}nnow$^{2}$
\\
$^{1}$Departamento de Astronomia, Universidad de Chile, Camino el Observatorio 1515, Las Condes, Santiago, Chile\\
$^{2}$Kapteyn Astronomical Institute, University of Groningen,Landleven 12, 9747 AD Groningen, The Netherlands\\
}

\date{Accepted XXX. Received YYY; in original form ZZZ}

\pubyear{2023}

\begin{document}
\label{firstpage}
\pagerange{\pageref{firstpage}--\pageref{lastpage}}
\maketitle

\begin{abstract}
Galactic halos accrete material from the intergalactic medium (IGM) and part of this accretion is expected to be in the form of cool ($T\sim10^4$ K) gas. A signature of this process could reside in the detection of numerous clouds in the circumgalactic medium (CGM). However, whether this material is able to accrete onto the galaxies and feed their star formation or, instead, evaporates into the CGM hot phase (corona, $T\sim10^6$ K), is not yet understood. Here, we investigate the evolution of cool CGM clouds accreted from the IGM and falling through the hot corona of low-redshift disc galaxies, using 3D high-resolution hydrodynamical simulations. We include the effects of gravity due to the dark matter halo, isotropic thermal conduction, radiative cooling and an ionizing UV background. We explored different values of parameters such as the halo mass, coronal mass, initial cloud velocity and strength of the thermal conduction. We find that the clouds lose the vast majority of their mass at distances larger than half of the galaxy virial radius and are completely dissolved in the corona before reaching the central galaxy. Resolving the Field length with at least 5-7 cells is crucial to correctly capture the evolution of the infalling cool gas. Our results indicate that cool IGM accretion can not feed star formation in $z\sim0$ star-forming galaxies in halos with masses of $10^{11.9}\ M_{\odot}$ or above. This suggests that present-day massive star-forming galaxies can sustain their star formation only via the spontaneous or induced cooling of their hot corona.
\end{abstract}

\begin{keywords}
galaxies:evolution -- galaxies:haloes -- hydrodynamics
\end{keywords}



\section{Introduction}\label{intro}
The standard cosmological model (Lambda cold dark matter, or $\Lambda$CDM) predicts that galaxy halos continuously grow due to accretion from the Cosmic Web. The growth of dark matter halos is accompanied by accretion of ordinary or \textit{baryonic} matter from the intergalactic medium (IGM), part of which eventually leads to the formation and growth of the central galaxy. However, while the accretion of collisionless dark matter into the halo is only governed by gravity and is well constrained in the $\Lambda$CDM framework, the fate of the collisional baryonic component, after it enters the halo and becomes part of the circumgalactic medium (CGM), is more complex and still not completely understood.

Accretion of gas is expected to happen in two main ways, or modes \citep{birnboim03}, which depend on the mass of the galactic halo, even though the exact demarcation between the two is still debated \citep[][]{fielding17,stern20a}. For galaxies at low redshift and that are massive enough, the gas is expected to accrete through a hot-mode: it is first shock-heated at high temperatures, forming what is known as galactic \textit{corona}, which will later cool and accrete into the center \citep[see also][]{rees77,white78}. This hot ($T\sim10^{6-7}$ K) corona has been observed at low redshift around massive galaxies through X-ray observations \citep[e.g.][]{bregman18,nicastro23}. Below some halo mass threshold, of the order of $10^{12} M_{\sun}$, cold-mode accretion is instead expected to be more common, with cold gas directly accreting towards the center, without ever reaching higher temperatures \citep[see also][]{binney77}. The two types of accretion can coexist with each other, with cold filaments penetrating the hot corona \citep[e.g.][]{dekel06,keres09}, although there is to date no consensus on whether these gas streams are able to accrete all the way into the central galaxy \citep[][]{dekel09} or are instead destroyed by the interactions with the hot gas \citep[][]{keres12,nelson13, nelson16,zheng20}.

A possible observational signature of this cool gas accretion resides in the cool ($T\sim10^{4}$ K) phase of the CGM, which has been detected both at high and low redshift \citep[see][and references therein]{farina19,tumlinson17}, including around massive galaxies which are known to be surrounded by a hot corona \citep[e.g.][]{werk13}. In particular, observations in absorption at low redshift have shown that the cool CGM is composed of multiple kinematically distinct components, often described as gas `clouds', which are located at hundreds of kpc from the center and exhibit a complex kinematics \citep[see][]{borthakur15, keeney17} and a wide range of metallicities, implying that at least part of them is coming from IGM accretion \citep[see][]{wotta19,afruni22}. It is possible that the cool circumgalactic clouds detected in the halos of massive galaxies originate from the fragmentation of accreting filaments, as a consequence of hydrodynamical instabilities that are expected to develop at the interface between the different gas phases \citep[e.g.][]{mandelker16}.

Ideally, a complete treatment of the origin and fate of these structures would require cosmological simulations at a sufficiently high resolution. Currently, however, despite the great improvements in the recent years, the resolution of large-box and even 'zoom-in' cosmological simulations \citep[e.g.][]{crain15,pillepich18,hopkins18} is still at best of the order of the size of 1 kpc in the CGM \citep[e.g.][]{peeples19,vandevoort19}, still significantly larger than the expected size of the cool gas structures \citep[e.g.][]{mccourt18}.  
Therefore, a crucial role remains to be played by idealized high-resolution (pc-scale) simulations, which are however necessarily focused on a relatively small portion of the galactic halo. Several of these studies have focused in the past on filaments moving through a hot medium \citep[e.g.][]{mandelker18,vossberg19,mandelker20}, or gas clouds moving in the proximity of galaxy discs \citep[e.g.][]{armillotta16,gronke18,kooij21}. In this work, we use high-resolution simulations to study cool CGM clouds accreted from the intergalactic medium at large distances from the central galaxy, aiming to understand whether they are able to reach the galaxy and feed its star formation.

A fundamental physical effect that needs to be considered when investigating the survival of cool clouds flowing through a hotter medium is thermal conduction, the transfer of heat via free electrons, which takes place in the presence of strong temperature gradients \citep[][]{spitzer62}. Analytical studies \citep[e.g.][]{nipoti07} have shown that thermal conduction tends to make cool clouds evaporate into the hot corona of massive passive galaxies, preventing cool material to accrete all the way to the center. On the other hand, being a diffusive process, thermal conduction also tends to hinder the development of the Kelvin-Helmoltz (KH) and Rayleigh-Taylor (RT) instabilities at the cloud/corona interface, which are also responsible for the cloud fragmentation \citep[e.g.][]{heitsch09,schneider17}, and can therefore extend the survival time of the cloud, as seen in several high-resolution simulations \citep[e.g.][]{vieser07,bruggen16,armillotta17}. One further complication is due to the effects of magnetic fields \citep[e.g.][]{gronnow18,kooij21}, which suppress the efficiency of thermal conduction, since electrons follow the magnetic field lines and can not move in the direction perpendicular to them. This is generally treated by introducing a suppression factor of the thermal conduction of the order of 0.1 \citep[see][and references therein]{kooij21}.

Instead of evaporating into the hot corona, an infalling cloud could also grow its cool mass through the condensation of the hot surrounding medium, an effect driven by the gas radiative cooling. The cool material stripped from the cloud creates gas at the cloud/corona interface at intermediate temperatures ($T\sim10^5\ \rm{K}$, close to the peak of the cooling function in collisional ionization equilibrium) and, depending on the initial conditions of the simulations (in particular the density of the corona), with short cooling times. As a result, the mixed gas cools very rapidly causing an increase of the mass of the cool medium \citep[see][]{fielding20}. This effect has been observed in several high-resolution simulations of the evolution of cool CGM clouds 
\citep[e.g.][]{marinacci10accr,armillotta16,gronke18,kooij21,tan23}. 
These studies are, however, focused on regions up to $\sim10\ \rm{kpc}$ from the galactic disc, at the interface between the disc and the CGM, where the density of the corona is relatively high, favouring the cooling of the mixed gas. One important exception was \citet{armillotta17}, who considered larger distances from the disc and lower gas densities, but ran their simulations in a 2D geometry, for a relatively short time (250 Myr), and did not study the effect of gravitational acceleration on the evolution of the gas clouds. In this work we aim, instead, to characterize the survival of the cool clouds accreting from the IGM and detected at hundreds of kpc from the galactic disc, following their journey from the virial radius to potentially the central galaxy.

We use high-resolution 3D hydrodynamical simulations with an adaptive grid (see Section~\ref{Simulations}), modelling the accretion of IGM cool clouds through the haloes and hot coronae of star-forming disc galaxies at low redshift, including halos resembling those of the Milky Way (MW) and M31. We use  analytical and observational arguments to define the initial conditions of the simulated CGM. 
In our simulations, we consider the effect of the gravitational field of the dark matter halo, which stratifies the corona and pulls the cloud inwards. We include radiative cooling, an ionizing UV background \citep{haardt12} and isotropic thermal conduction, exploring different levels of suppression due to magnetic fields. Thanks to the adaptive grid, we are able to properly resolve the cool gas (our maximum resolution is smaller than the Field length, which is the critical physical scale for evaporation), and at the same time to follow the cloud infall from the virial radius towards the central galaxy. The main goal of this study is to determine whether cool clouds accreting from the IGM can reach the central galaxy and feed star formation therein, or are instead destroyed by the interactions with the hot corona.

In Section~\ref{Simulations} we describe in detail the initial conditions and the physical processes included in our hydrodynamical simulations; in Section~\ref{results}, we report the results of our numerical experiments; in Section \ref{discussion}, we discuss the limitations of this work, the comparison with previous studies and the implications of our findings for our understanding of galaxy evolution. Finally, in Section~\ref{conclusions}, we summarize our work and we outline our conclusions.

\section{Hydrodynamical simulations}\label{Simulations}
In this Section, we report the initial conditions and the physical processes of the high-resolution hydrodynamical simulations performed in this study. To carry out our numerical experiments, we used the code PLUTO, version 4.3 \citep[see][]{mignone07,mignone12}, which is an Eulerian Godunov-type \citep{godunov59} code that solves the system of hydrodynamical equations. We used in particular the HLLC Riemann solver \citep[][]{toro94}, which represents a good compromise between accuracy and stability of the simulations \citep[see][]{gronnow18}.

\subsection{Initial conditions}\label{section2}
\subsubsection{Dark matter halos}
We focused on three different galactic halos representative of disc galaxies of different masses in the Local Universe. We investigated halos of virial masses ($M_{\rm{vir}}$), equal to  $10^{11.9}, 10^{12.1}$ and $10^{12.3}\ M_{\odot}$ and virial radii ($r_{\rm{vir}}$) equal to $252, 294$ and $336$ kpc.  
Note that the high- and intermediate- mass halos are consistent with respectively those of M31 \citep[see][]{afruni22} and the Milky Way \citep[e.g.][]{postiHelmi19}, hence hereafter we will refer to them as M31 and MW halos, while we will call the lowest-mass halo Sub-MW. Using the stellar-to-halo-mass relation derived in \cite{posti19b}, these virial masses correspond to stellar masses equal to $10^{10.1}, 10^{10.4}$ and $10^{10.7}\ M_{\odot}$, corresponding to the median stellar masses of the three bins considered in \cite{afruni20}, who studied the cool CGM of a sample of star-forming galaxies at low redshift. 
The dark matter distribution follows the Navarro Frenk White profile \citep[NFW][]{nfw96}, assuming the corresponding virial mass and radius and a concentration $c$ calculated from \cite{dutton14}. 

\subsubsection{Density and temperature of the hot CGM}\label{hotmedium}
For each of the three considered halos, we define a hot corona as a static medium in hydrostatic equilibrium within the gravitational potential of the dark matter. In particular, we assume a polytropic hydrostatic equilibrium, in which the density and temperature are described by
\begin{ceqn}
\begin{equation}\label{eq:corn}
n_{\rm{cor}}(z)=n_0\left(\frac{T_{\rm{cor}}(z)}{T_{\rm{cor},0}} \right)^{1/(\gamma-1)}\ ,
\end{equation}
\end{ceqn}
and
\begin{ceqn}
\begin{equation}\label{eq:corT}
\frac{T_{\rm{cor}}(z)}{T_{\rm{cor},0}}=1+\frac{\gamma-1}{\gamma}\frac{\mu m_{\rm{p}}}{k_{\rm{B}} T_{\rm{cor},0}}\left(\Phi(z)-\Phi_0 \right)\ ,
\end{equation}
\end{ceqn}\\
where $\mu\simeq0.6$ is the mean molecular weight (see also Section~\ref{radprocess}), $m_{\rm{p}}$ is the proton mass, $k_{\rm{B}}$ is the Boltzmann constant, $\gamma=1.2$ is the polytropic index, $T_{\rm{cor},0}$, $n_0$ and $\Phi_0$ are the temperature, density and gravitational potential at a reference radius $r_0 = r_{\rm{s}}$, where $r_{\rm{s}}$ is the scale radius (equal to $r_{\rm{vir}}/c$) and $\Phi(z)$ is the NFW potential. We assume the gravitational potential to be given only by the DM component, neglecting the effects of the stellar disc and possibly bulge. At the large distances probed with our simulations, the contribution of the stellar components to the total potential is indeed negligible. Note that we made the assumption that the density and temperature of the corona are plane parallel, hence they vary only as a function of the height $z$. Neglecting curvature effects is justified as long as the gas belonging to a single cool cloud always maintains a transverse extent that is much smaller than the distance from the galaxy, a condition that is always met in our simulations.
   \begin{figure}
   \includegraphics[clip, trim={0cm 0cm 0cm 0cm}, width=\linewidth]{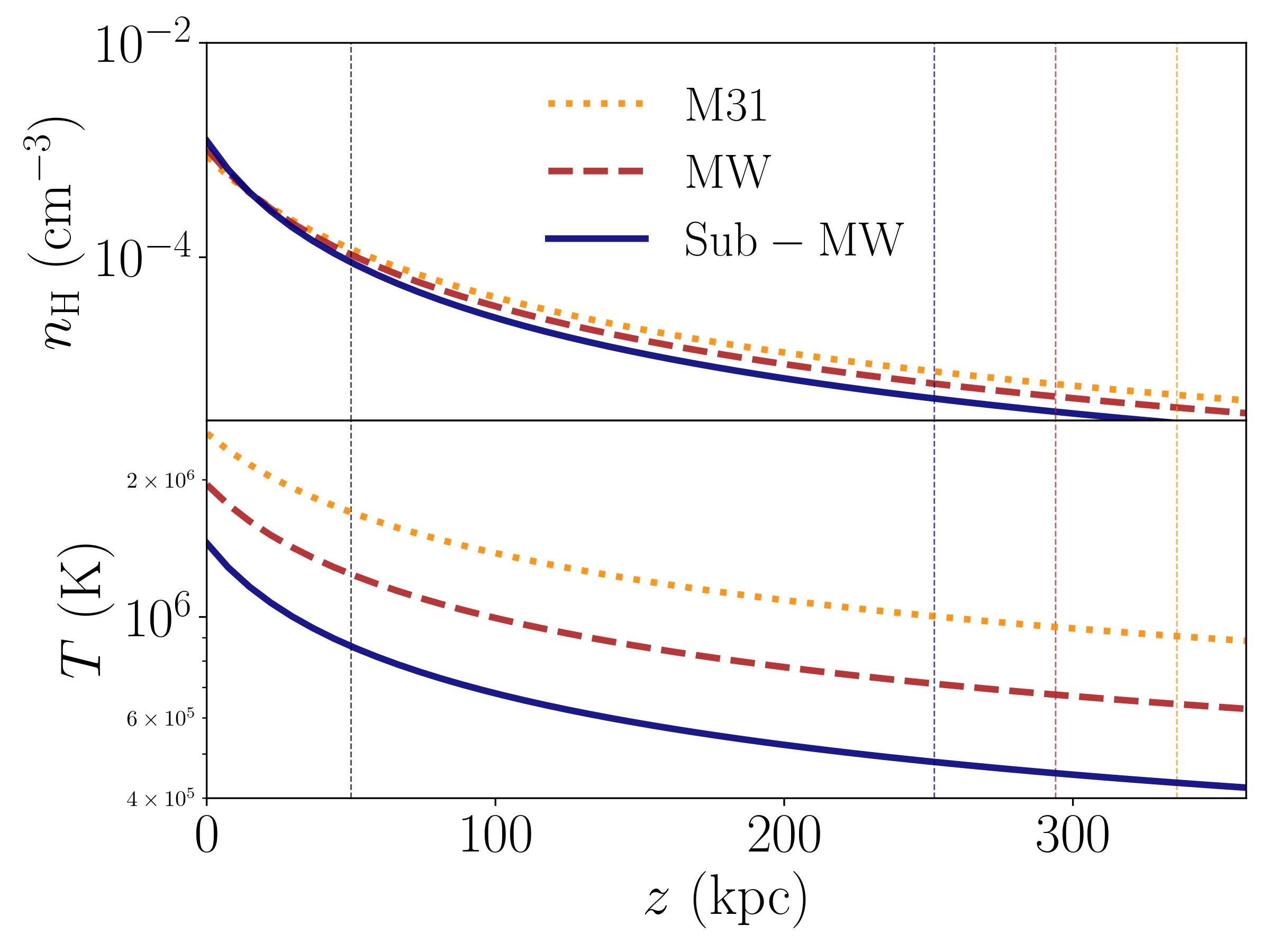}
   \caption{Profiles of hydrogen number density (top) and temperature (bottom) of the hot corona in our simulations as a function of the distance from the galactic disc, derived respectively from equations~\eqref{eq:corn} and \eqref{eq:corT}, for the three halos considered in this work, assuming that the corona contains 20\% of the theoretical baryonic budget within the virial radius $r_{\rm vir}$ (indicated for each halo mass by a colored vertical dashed line). Our simulation box extends from $z_{\rm min} =50 \; {\rm kpc}$ (indicated by the black vertical dashed line) to $z_{\rm max} = r_{\rm vir} + 10 \; {\rm kpc}$.}
              \label{fig:Hotgas}%
    \end{figure}
In order to simplify the setup of the simulations, we further assume that the corona does not rotate (see Sections~\ref{limitations} for more details). We chose values of $\gamma$, $T_{\rm{cor},0}$ and $n_0$ that produce a model of the hot gas whose density and temperature profiles are consistent with those of the more sophisticated, rotating hot CGM model presented in \cite{afruni22} for M31.

In most of our simulations, the total coronal mass is equal to 20\% of the total baryonic mass expected within the halo ($M_{\rm{vir}}\times f_{\rm{bar}}$, where $f_{\rm{bar}}=0.158$ is the cosmological baryon fraction, \citealt{planck20}). This choice is justified by the fact that the corresponding density profiles have values consistent with the available observational estimates \citep[e.\ g.][]{gatto13,salem15,putman21}. The density and temperature profiles of the corona for the three halos analyzed in (most of) our simulations are shown in the two panels of Figure~\ref{fig:Hotgas}. In some simulations we have assumed a coronal mass within the virial radius equal to 40\% of the baryonic budget (see Table~\ref{tab:inCond}). In these cases, we adopt a density profile equal to the one shown in the top panel of Figure~\ref{fig:Hotgas} multiplied by a factor of 2. In all simulations, we adopted a metallicity $Z=0.3\ Z_{\odot}$ everywhere in the hot corona, consistently with observations of the inner regions of the corona of the Milky Way and nearby massive spiral galaxies \citep[see][]{miller15,anderson16}. We note that the metallicity of the corona, at distances of hundreds of kpc from the disc, is less well constrained and may be lower than our adopted value. We will discuss the impact of this choice on our results in Section~\ref{limitations}.

\subsubsection{Properties of the cool clouds}\label{coolCloudsProps}
Once the properties of the hot CGM are defined, we assume that a cool ($T = 2\times10^4$ K, see \citealt{keeney17,lehner20p}) cloud, pressure-confined by the hot corona, is infalling from the virial radius towards the galactic disc. We assume a metallicity in the cloud $Z=0.05Z_{\odot}$, which is consistent with observational estimates \citep[][]{lehner18} and with the results of the semi-analytical models of \cite{afruni22}.

To fully define the initial conditions, we need to specify an initial cloud mass and initial velocity. Note that, given the assumption of initial pressure equilibrium, the radius of the cloud is uniquely defined by its mass. Cloud mass and initial velocity are poorly constrained, as they depend on the fragmentation of the IGM filaments when impacting the pre-existing hot CGM, a process which is to date not well understood (see also Section~\ref{limitations}). We therefore considered different values of the initial velocity of the cloud and we then estimated plausible associated values for the mass using the theoretical and observational arguments explained in the following.

In the scenario of spherically symmetric gas accretion, the flux of mass at a radius $r$ is described by:
\begin{ceqn}
\begin{equation}\label{eq:faccr}
F_{\rm{accr}} = 4 \pi r^2 n_{\rm{cl}} m_{\rm{cl}} v_{\rm{cl}}\ ,
\end{equation}
\end{ceqn}
where $n_{\rm{cl}}$ is the number of clouds per unit of volume, $m_{\rm{cl}}$ is the cloud mass and $v_{\rm{cl}}$ is the cloud velocity. We can then define the covering fraction (or factor) of the cool circumgalactic gas at a projected radius $R$ as:
\begin{ceqn}
\begin{equation}\label{eq:covfrac}
f_{\rm{c}} = \pi r_{\rm{cl}}^2 n_{\rm{cl}} \langle d \rangle\ ,
\end{equation}
\end{ceqn}
where $\langle d \rangle\ = 2 \sqrt{r_{\rm{vir}}^2 - R^2}$ is the extension of the CGM along the line of sight, considering clouds distributed within the virial radius, and $r_{\rm{cl}}$ is the cloud radius, defined as:
\begin{ceqn}
\begin{equation}\label{eq:mcl}
r_{\rm{cl}} = \left(\frac{3}{4\pi \rho_{\rm{cl}}} m_{\rm{cl}}\right)^{1/3}\ ,
\end{equation}
\end{ceqn}
where $\rho_{\rm{cl}}$
is the density of the cloud, calculated from the assumption of pressure equilibrium with the hot corona. Combining the previous equations and solving for $n_{\rm{cl}}$, we obtained an expression for the cloud mass that depends on its velocity, on the total flux of mass accretion and on the observed covering fraction:
\begin{ceqn}
\begin{equation}\label{eq:mmax}
m_{\rm{cl}} = \left ( a \frac{F_{\rm{accr}}}{ 4r_{\rm{vir}}v_{\rm{cl}}f_{\rm{c}}} \right) ^ 3 \left(\frac{3} {4\pi \rho_{\rm{cl}}}\right)^2,
\end{equation}
\end{ceqn}
where $a = \langle d \rangle\ / r_{\rm{vir}}$. The flux of gas accretion at the virial radius depends on the virial mass of the host galaxy and we estimated it for each halo using the prescriptions from the cosmological models of \cite{correa15a,correa15b}.
 
We can now connect this framework with observational constraints on the covering factor of cool CGM clouds $f_{\rm c}$ in the proximity of the virial radius (i.e.\ where our initial conditions are to be defined), as derived from absorption studies (see Section \ref{intro}). Note that, for application to equation~\eqref{eq:covfrac}, the relevant definition of covering factor is the average number of clouds per line of sight, which is allowed to be larger than one, as multiple clouds (or `components') can (and are often observed to) overlap along the same line of sight. This quantity is however not always easy to measure in observations (because it requires sufficient spectral resolution and sometimes involves subjective choices in the separation of the components). Here we adopt the simpler and more common definition of $f_{\rm c}$ as a `detection frequency', or fraction of lines of sight with a detection. This by definition cannot be larger than one and should be regarded as a lower limit to the actual covering factor. We will argue in Section \ref{limitations} that this is a conservative choice and adopting a more proper definition of $f_{\rm c}$ would only go in the direction to reinforce the conclusions of this paper. With these considerations in mind, we estimated the covering fraction from observations of the cool CGM around low-redshift star-forming galaxies with masses similar to those analyzed in this work, by selecting all the sightlines with $0.7r_{\rm{vir}}<R<r_{\rm{vir}}$ in the sample from \cite{lehner20p} for the M31 halo and from \cite{borthakur15} for the MW and Sub-MW halos (adopting the same mass binning as in \cite{afruni20}). We obtained covering fractions $f_{\rm{c}}$ for the M31, MW and Sub-MW samples of respectively 0.86, 0.83 and 1.
These values are also in broad agreement with independent estimates in the recent literature \citep[e.g.][]{wilde21}.

Once the covering fraction was estimated, we calculated the quantities $a, v_{\rm{cl}}$ and $\rho_{\rm{cl}}$ at a radius $r_{\rm{med}}$, defined as the center of the considered radial bin, and found $m_{\rm cl}$ from equation \eqref{eq:mmax} for a given value of the initial velocity. The cloud density $\rho_{\rm cl}$ was obtained under the assumption of pressure equilibrium with the hot gas and is therefore sensitive to the assumed properties of the corona, in particular its total mass. We took into account that $v_{\rm{cl}}(r_{\rm{med}})$ is not exactly the same as the initial velocity, due to non-negligible acceleration in the considered radial bin.\footnote{In detail, we calculated a velocity profile using an approximated expression for the drag force exerted on the cloud by the hot corona \citep[see][]{afruni19}. We have verified a posteriori that this approximation is a very good description of our simulations in the radial range of interest (see Section \ref{implications:absorbers}). Because the drag acceleration depends on the cloud mass, this calculation required an iterative approach, which converged in a few steps.}
Note that, given equation~\eqref{eq:mmax}, higher-velocity clouds, as well as clouds embedded in more massive coronae (and therefore denser), need to be less massive, in order to remain consistent with the observational constraints on the covering factor. We will come back to this point in Sections \ref{ResCoronalMass} and \ref{resVelocity}.

To define the initial conditions of the cool clouds in our simulations (listed in Table~\ref{tab:inCond}), we chose, for each halo, two different values for the initial infall velocity (10 and 100 km s$^{-1}$) and for the mass of the hot corona (20 and 40\% of the total baryonic mass), and for each combination we derived an associated value of the initial cloud mass according to equation \eqref{eq:mmax} following the method described above. The only exception was made for the M31 simulations where $v_{\rm{cl}}(r_{\rm{vir}})= 10$ km s$^{-1}$ and $f_{\rm{cor}}=M_{\rm{cor}}/M_{\rm{bar}}=0.2$. In these cases, we used the cloud mass coming directly from the findings of the Bayesian analysis of \cite{afruni22}. In that study, the authors built semi-analytical models of gas accretion into the halo of the Andromeda galaxy and compared the results with the state-of-the-art observations of the AMIGA project \citep[][]{lehner20p}. They found an initial cloud mass of $5\times10^6\ M_{\odot}$, for an initial infall velocity of about 10 km s$^{-1}$. Note that using equation~\eqref{eq:mmax}, we would obtain a slightly smaller (a factor 2) value than what found by \cite{afruni22}. This might be due to the fact that in the latter the total mass accretion was slightly higher than the cosmological value, which we are using in our current calculations. Given the higher accuracy of the models of \cite{afruni22} with respect to the simple analytical calculations described above, we decided to adopt their mass value for these specific simulations.

Once the initial mass (and therefore the radius $r_{\rm{cl}}$) of the cloud is defined, the density profile at the interface between the cloud and the corona is defined as a smooth function \citep[see][]{gronnow18,kooij21}:
\begin{ceqn}
\begin{equation}\label{eq:shellChap5}
n(r)=n_{\rm{cor}} + 0.5(n_{\rm{cl}}-n_{\rm{cor}})\left\{1 - \tanh\left[s(r/r_{\rm{cl}}-1) \right] \right\} ,
\end{equation}
\end{ceqn}
where the parameter $s$ sets the steepness of the profile (we adopted $s=10$ for all our simulations). This smoothed top-hat profile produces a smooth transition between the coronal and the cloud density, with $n(r_{\rm{cl}})\approx n_{\rm{cl}}/2$ in the limit where the density of the cloud is significantly higher than the one of the corona, which is the case in all our experiments.\footnote{Note that the transition profile is quite steep, therefore the cloud initial properties are still approximately constant throughout most of the cloud. Nonetheless, the initial mass of the cloud will be slightly ($\sim5\%$) smaller than the nominal value reported in Table~\ref{tab:inCond}.} Note that, as a consequence of pressure equilibrium, the interface between the cloud and the corona is also characterized by a smooth temperature profile that varies from $2\times10^4\ \rm{K}$ to the temperature of the corona. 

{ 
 \begin{table*}
\begin{center}
\begin{tabular}{*{9}{c}}
(1)&(2)&(3)&(4)&(5)&(6)&(7)&(8)&(9)\\
\hline  
\hline
 Sim. Id & $\log \frac{m_{\rm{cl}}}{M_{\odot}}$ & $r_{\rm{cl}}$ & $v_{\rm{cl}}$ & $f_{\rm{tc}}$ & $f_{\rm{cor}}$ & $\log \frac{M_{\rm{halo}}}{M_{\odot}}$ & $r_{\rm{vir}}$ & Max res. \\
     & & (kpc)  & (km s$^{-1}$)  &   &  &  & (kpc) & (pc)\\
\hline 
M31FID & $6.7$ & 5.2 & $10$ & $0.1$& $0.2$ & 12.3& 336 & $62.5$\\
M31F001 &$6.7$ & 5.2 & $10$& $0.01$& $0.2$ & 12.3& 336 & $62.5$\\
M31B40 & $5.86$ & 2.2 & $10$& $0.1$& $0.4$ & 12.3& 336 & $62.5$\\
M31100 & $6.00$ & 3.1 & $100$& $0.1$& $0.2$ & 12.3& 336 & $62.5$\\
MWFID & $6.48$ & 5.0 & $10$& $0.1$& $0.2$ & 12.1& 294 &$62.5$\\
MWHRES & $6.48$ & 5.0 & $10$& $0.1$& $0.2$ & 12.1& 294 & $31.25$\\
MWLRES & $6.48$ & 5.0 & $10$& $0.1$& $0.2$ & 12.1& 294 & $125$\\
MWF001 & $6.48$ & 5.0 & $10$& $0.01$& $0.2$ & 12.1& 294 & $62.5$\\
MWB40 & $5.98$ & 2.7 & $10$& $0.1$& $0.4$ & 12.1& 294 & $62.5$\\
MW100 & $6.05$ & 3.6 & $100$& $0.1$& $0.2$ & 12.1& 294 & $62.5$\\
SMWFID & $6.39$ & 5.3 & $10$& $0.1$& $0.2$ & 11.9& 252 & $62.5$\\
SMWF001 & $6.39$ & 5.3 & $10$& $0.01$& $0.2$ & 11.9& 252 & $62.5$\\
SMWB40 & $5.90$ & 2.9 &$10$& $0.1$& $0.4$ & 11.9& 252 & $62.5$\\
SMW100 & $5.89$ & 3.6 & $100$& $0.1$& $0.2$ & 11.9& 252 & $62.5$\\
\hline
\end{tabular}
\end{center}
\captionsetup{}
\caption[]{List of all the simulations performed in this study, with the parameters that define the initial conditions: (1) Id chosen for the simulation; (2) initial cloud mass; (3) initial cloud radius; (4) initial infall velocity of the cloud; (5) suppression factor of the thermal conduction; (6) fraction of baryonic mass present in the hot corona; (7) mass of the DM halo; (8) virial radius, which is also the starting point for our simulated infalling clouds; (9) maximum grid resolution. The naming scheme adopted for the simulation Ids is the following: M31, MW and SMW are used to identify the halo mass and FID is used for the fiducial simulations (see Section~\ref{setup}); the rest of the Id emphasises variations from the fiducial configuration, in terms of the resolution (HRES/LRES), the suppression factor (F001), the coronal mass (B40), and the initial cloud velocity (100).}\label{tab:inCond}
   \end{table*}
   }

\subsection{Radiative processes and thermal conduction}\label{radprocess}
We include, in our simulations, the effects of radiative cooling, an ionizing extragalactic UV background and of thermal conduction. The cooling and heating rates (respectively $\Lambda$ and $\Gamma$) and the mean molecular weight $\mu$ are calculated through the code CLOUDY \citep[version c13; ][]{ferland13}, using the ionizing UV background of \cite{haardt12} at $z=0$ and the abundances from \cite{sutherland93}. More in detail, the cooling and heating rates and the values of the mean molecular weight are tabulated as a function of the temperature $T$, the metallicity $Z$ and the hydrogen density $n_{\rm{H}}$, in a similar fashion as in \cite{armillotta17}, to which we refer for more details. In our tables the temperature varies from $10^3$ to $10^7$ K, the metallicity from $0.01Z_{\odot}$ to the solar value and the hydrogen density from $10^{-6}$ to $1$ cm$^{-3}$.
The total net energy loss/gain per unit time and volume due to cooling and heating is given by $\Delta E=-n^2\Lambda_{\rm{net}}(T,Z,n_{\rm{H}})$, where $\Lambda_{\rm{net}}=\Lambda-\Gamma$.

As already mentioned, in all the simulations the cloud has a metallicity $Z=0.05\,Z_{\odot}$, while the corona has $Z=0.3\, Z_{\odot}$. We keep track of the metallicity in the simulations in the same way as in \cite{kooij21}. We assume that the corona is in thermal equilibrium, likely due to the presence of heating sources (for example from the central galaxy) that balance cooling. For simplicity, we do not explicitly include such heating terms and instead we directly assume that the hot gas can not cool. In practice, we set a passive tracer $C_{\rm{cool}}$ (see \citealt{mignone12} for the evolution of passive scalars in PLUTO), which is used to isolate the coronal material that is not affected by the cloud. This tracer is initially set to 1 for the cloud (for $r<1.3r_{\rm{cl}}$, in order to contain all the cloud material) and 0 elsewhere and evolves with the simulation. At anytime, material where $C_{\rm{cool}}<10^{-8}$ (indicating purely coronal gas, i.e.\ not mixed with gas that was originally in the cloud) is not allowed to cool. We note that this choice will affect primarily the innermost parts of the corona, which may otherwise cool significantly during the simulation, given the higher density and the long running times.

In the simulations, thermal conduction is introduced by adding an additional divergence term in the energy equation, using the module available in PLUTO \citep[see][]{mignone12}. The flux of thermal conduction can be in the non-saturated or saturated regime, depending on whether the mean free path of electrons is respectively smaller or larger than the temperature scale length \citep[see][and references therein, for more details]{armillotta17,kooij21}.
In the non-saturated regime, we multiply the Spitzer standard flux \citep{spitzer62} by a suppression factor $f_{\rm{tc}} < 1$ to effectively take into account the suppression of thermal conduction due to the magnetic field. Note that we are neglecting, for simplicity, anisotropic effects that would in reality be present due to the fact that heat conduction is allowed along magnetic field lines (along which electrons can freely move) and suppressed in perpendicular directions. This isotropic approximation is justified, on average, in turbulent flows (such as those developing as a consequence of hydrodynamical instabilities) as the magnetic field in this case tends to not have a well defined direction. 
Recently, the effective (isotropic) suppression factor has been estimated for a CGM cloud/corona system with MW-like properties, by comparing magneto-hydrodynamic (MHD) high-resolution simulations with anisotropic thermal conduction with hydrodynamic simulations including isotropic thermal conduction, by \cite{kooij21}. These authors constrained the value of $f_{\rm{tc}}$ to be between 0.03 and 0.15.

\subsection{Simulation setup}\label{setup}
All the simulations in this study are in 3D and the simulation box extends from $-50$ to 50 kpc in the $x$ and $y$ directions and from 50 kpc to $r_{\rm{vir}}+10$ kpc in the $z$ direction, with the cloud starting its infall at $x = 0, y=0, z=r_{\rm{vir}}$ and gravity pointing to $z=0$.

Our numerical experiments have been performed using the Adaptive Mesh Refinement (AMR) technique, provided by the PLUTO package. This is crucial in order to be able to follow the evolution of the cloud in a reasonable computational time. The cloud occupies only a small fraction of the grid, but might eventually travel for hundreds of kpc. Hence, the simulated volume must be large to capture the whole journey of the cloud, but only a fraction of it (where the cool gas is located) needs to be resolved at any one time.
In particular, the cells are refined according to a refinement threshold $\chi$. We utilized the default refinement criterion implemented in PLUTO: a cell is refined whenever the second derivative error norm \citep[see][]{lohner87} of the density exceeds the selected threshold. We have adopted, in all our simulations, $\chi=0.6$. We have performed several test runs using different choices of $\chi$ and we found that this value represents the best compromise in order to have the required resolution without being too computationally expensive ($\chi=0$ corresponds to refinement everywhere, while $\chi=1$ corresponds to no refinement). In order to capture the entire evolution of the cloud, we ran all our simulations for 3 Gyr.

The base grid of the simulation has a resolution of $1$ kpc and in most of our experiments we refine up to 4 levels, increasing the resolution by a factor 2 for each level, to a maximum resolution, therefore, of $62.5$ pc, which we consider our standard maximum resolution. We refine one level more and one level less in respectively the high- and low-resolution cases to test convergence. As can be seen in Table~\ref{tab:inCond}, the initial cloud radii vary from about $3$ to $5$ kpc and are therefore well above the grid resolution of the highest levels. This ensures that the cloud is in all cases very well resolved. We discuss more in detail the convergence of our simulations in Section~\ref{convergence}.

All the simulations that we have run for this work are reported in Table~\ref{tab:inCond}. We define one fiducial simulation setup for each one of the three halos analyzed (denoted with the 'FID' id). These are simulations where the coronal mass is equal to $20\%$ of the total baryonic mass (see Section~\ref{section2} and Figure~\ref{fig:Hotgas}), the suppression factor of the thermal conduction is equal to 0.1 (see the above considerations in Section~\ref{radprocess}) and the initial infall velocity of the cloud is equal to $10$ km s$^{-1}$, with the corresponding cloud mass coming from equation~\eqref{eq:mmax} or from \cite{afruni22} in the case of M31. We then performed more tests varying these three different parameters, $f_{\rm{cor}}, f_{\rm{tc}}$ and $ v_{\rm{cl}}(r_{\rm{vir}})$, to assess the robustness of our findings. 

\section{Results}\label{results}
In this section, we analyze the results of the simulations listed in Table~\ref{tab:inCond}. We start in Section \ref{subsec:fiducial} with the fiducial setups for each one of the three halos considered in this study (see Section~\ref{section2}), then we explore variations in the suppression factor of the thermal conduction (Section \ref{ResConduction}), total mass of the corona (Section \ref{ResCoronalMass}) and initial velocity of the cool cloud (Section \ref{resVelocity}) and finally present and discuss our tests of numerical convergence (Section \ref{convergence}). 

Throughout the analysis, we focus primarily on the survival of the cloud: to this end, we will show the evolution of the cool gas, as a function of time and distance from the central galaxy. In particular, we define an \textit{evaporation time} $t_{\rm{ev}}$ and an \textit{evaporation radius} $z_{\rm{ev}}$, as the time when and distance where the cloud has lost 90\% of its initial mass. These quantities give us an estimate of how long and how far can the cloud travel through the galactic halo. After this point we can consider the cloud effectively evaporated into the hot medium, as it will become evident from the results of our simulations shown below. In our main analysis, the cool gas is defined as the material with a temperature $T<3\times10^4$ K. Note that this threshold is relatively arbitrary, but our results do not depend on this choice, as we discuss more in  detail in Section~\ref{limitations}.

\subsection{Fiducial simulations}\label{subsec:fiducial}
   \begin{figure}
   \includegraphics[clip, trim={5.5cm 0cm 4.2cm 1cm}, width=\linewidth]{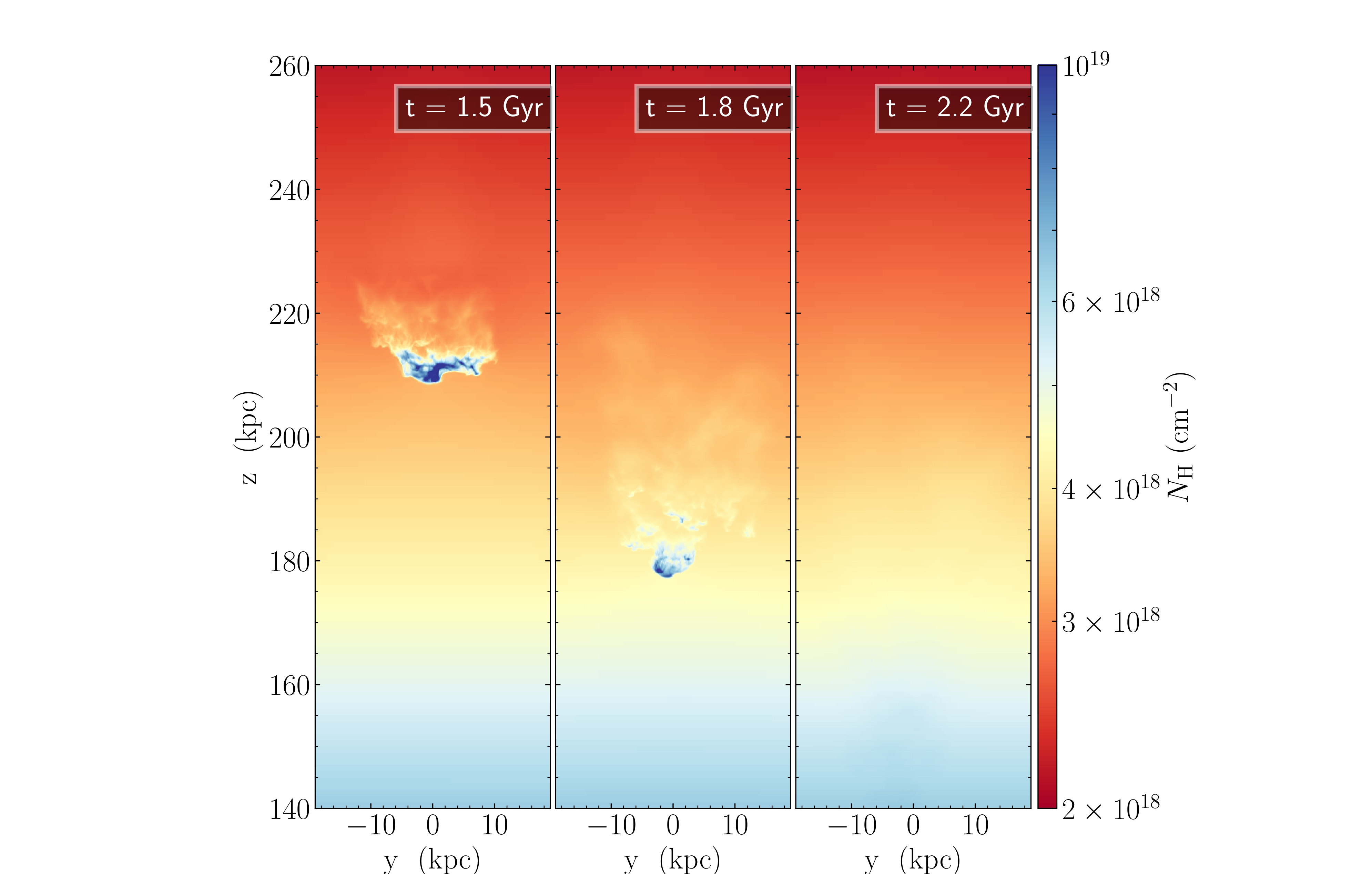}
   \includegraphics[clip, trim={5.5cm 0cm 4.2cm 1cm}, width=\linewidth]{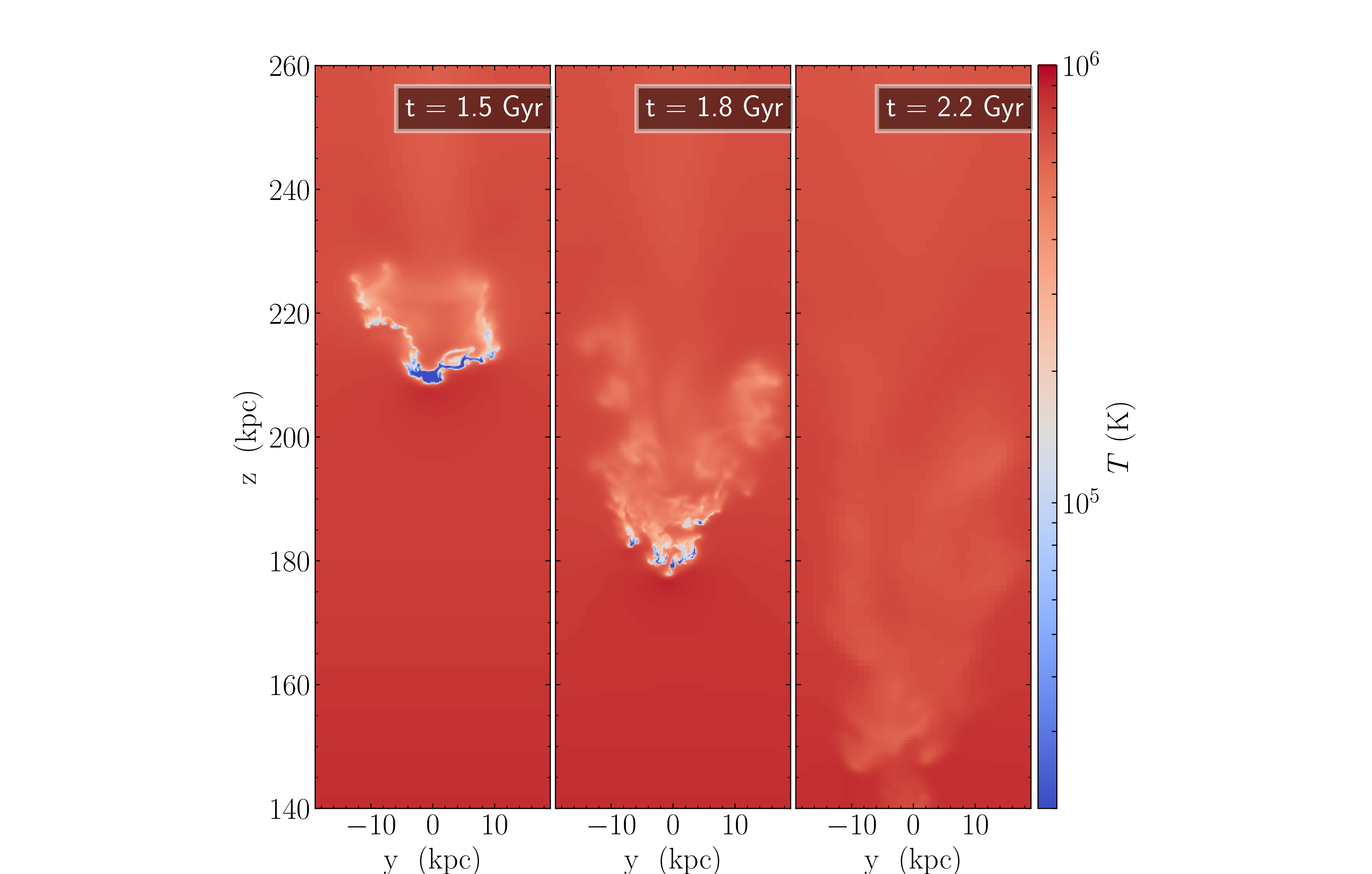}
   \caption{Total hydrogen column density projections along the $x$-axis (top) and temperature slices at $x=0$ (bottom), in a zoom-in of the simulation box that contains the cloud, at three different times (1.5, 1.8 and 2.2 Gyr), for the simulation MWFID.}
              \label{fig:snapshots}%
    \end{figure}
In Figure~\ref{fig:snapshots} we show, as a reference, selected density and temperature snapshots of the simulation MWFID, our fiducial simulation (see Section~\ref{coolCloudsProps}) of an IGM cloud infalling within the halo of a MW-like galaxy. The results for the fiducial M31 and SMW cases are comparable to what shown here. Both panels show three different snapshots at 1.5, 1.8 and 2.2 Gyr, with the top showing the total hydrogen column density, projected along the x-axis, and the bottom showing temperature slices at $x=0$. The cloud starts its infall at a height of almost 300 kpc (corresponding to the galaxy virial radius) and after 1.5 Gyr, accelerated by the gravitational force, has travelled down by about 100 kpc, almost reaching a distance of 200 kpc from the galactic disc. Even though it still looks relatively compact, at this point the cloud has completely lost its initial spherical shape. In particular, we can see that the cloud has been flattened in the $z$-direction due to the ram pressure exerted by the hot medium. Moreover, we can already note the onset of the Kelvin-Helmoltz and Rayleigh-Taylor instabilities. The development of both instabilities is more visible at 1.8 Gyr, where the cloud is clearly fragmented in multiple smaller cloudlets. Finally, at 2.2 Gyr, the cool gas is completely evaporated in the hot corona and is no longer present in the simulation box. This represents the main result of this study: the cool circumgalactic cloud, accreted from the IGM, is not able to reach the galactic disc and its fate is to be destroyed by the interactions with the hot surrounding CGM and join the hot corona. We refer to Section \ref{subsec:processes} for a discussion of the hydrodynamical processes that contribute to the cloud disruption and their associated time-scales.

\begin{figure}
   \includegraphics[clip, trim={0cm 0cm 0cm 0cm}, width=\linewidth]{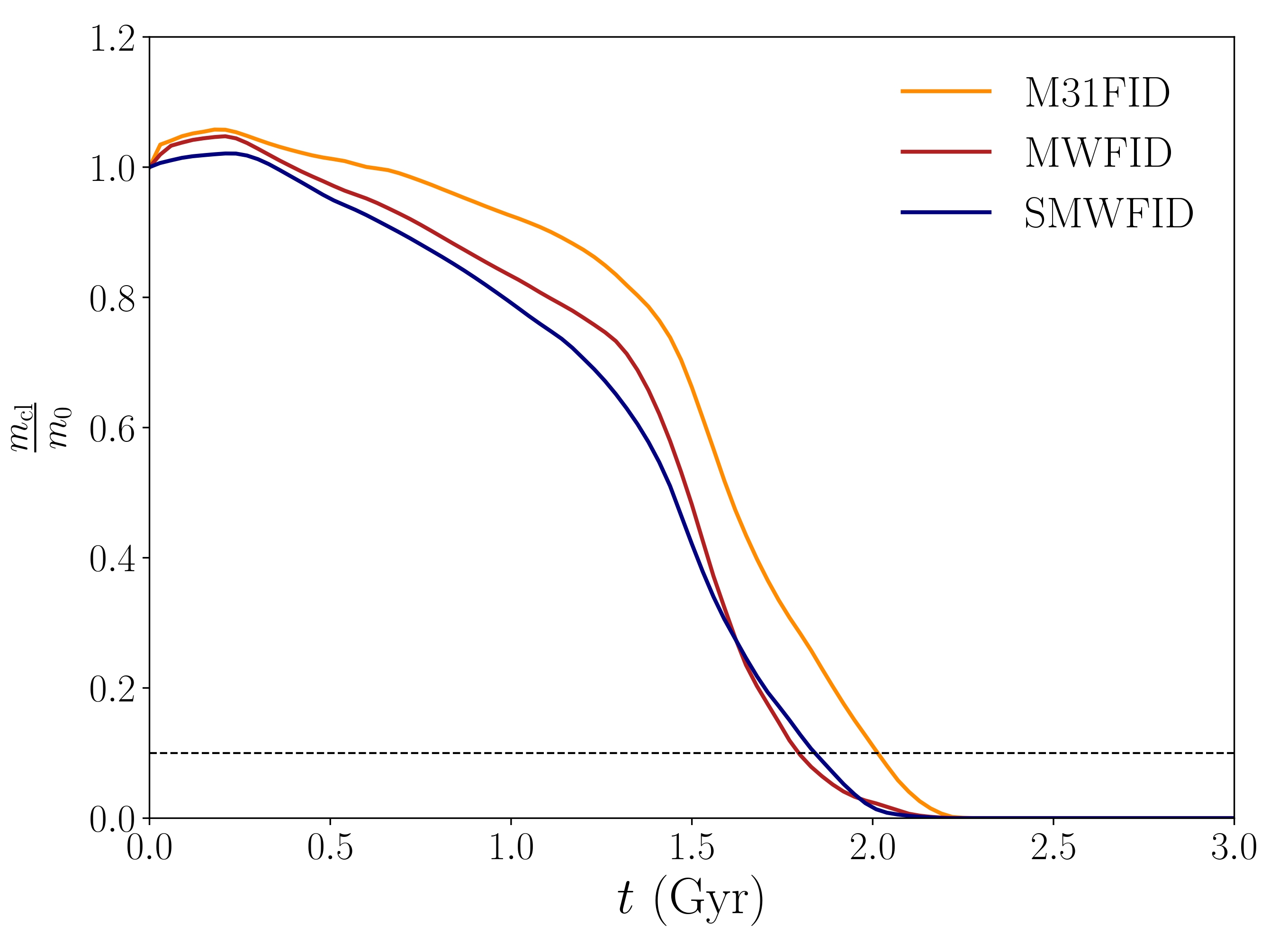}
   \includegraphics[clip, trim={0cm 0cm 0cm 0cm}, width=\linewidth]{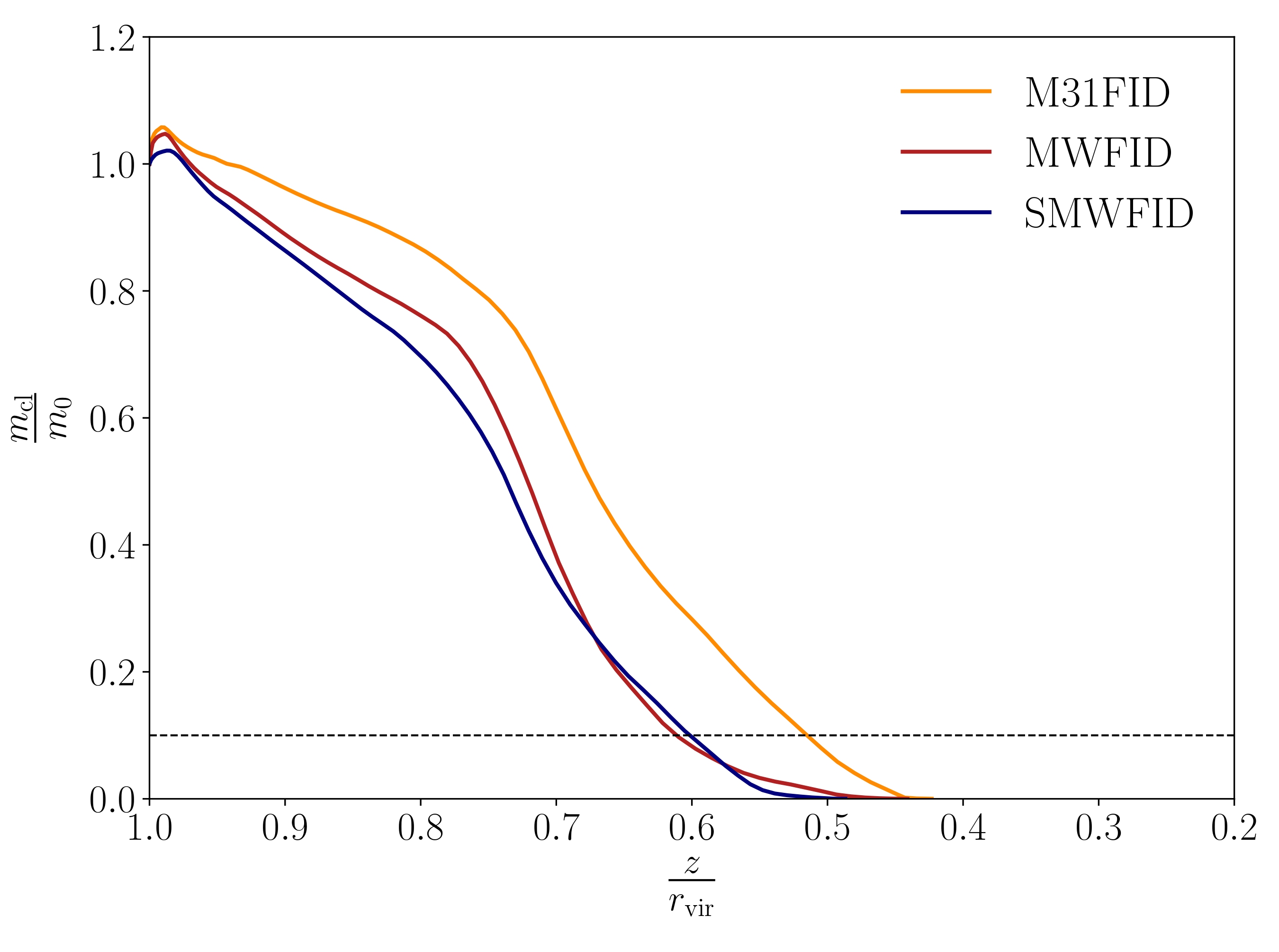}
   \caption{Evolution with time (top) and normalized distance from the central galaxy (bottom) of the cool ($T<3\times10^4$ K) gas mass present in the simulation box, normalized with the initial cool cloud mass, for our three fiducial simulations, M31FID, MWFID and SMWFID. The horizontal dotted line marks the level where the cloud has lost 90\% of its initial mass, i.e.\ $m_{\rm{cl}}/m_0=0.1$. In all our fiducial simulations, accreting IGM clouds are destroyed well before they can reach the central galaxy.}
              \label{fig:profilesFID}%
    \end{figure}

To be more quantitative, in the following we analyze the behavior of the cool gas as a function of time and distance from the disc. To this end, as mentioned above, we calculated at each timestep the mass of the cool medium, by selecting all the cells where the gas has a temperature $T<3\times10^4$ K. The two plots in Figure~\ref{fig:profilesFID} show the mass of cool gas, normalized to the initial cloud mass, as a function of time (top) and of the distance from the galaxy, reached during the infall, in units of the virial radius (bottom). The three different curves show the results of the three fiducial simulations, MWFID (red), M31FID (orange) and SMWFID (blue), while the horizontal line marks the level where the cloud has lost 90\% of its initial mass.

We can note how the three profiles exhibit a similar behavior: after a small initial condensation, due to the cooling of the shell of gas at intermediate temperatures that we set in the initial conditions between the cloud and the hot corona (see Section~\ref{coolCloudsProps}), the cloud starts losing cool gas, at a slow rate first, and more rapidly after about 1.5 Gyr (at $\sim0.7\ z/r_{\rm{vir}}$). In all cases, the cool gas completely evaporates into the corona within 2.5 Gyr, before having reached a distance from the disc of 0.4$r_{\rm{vir}}$. The cloud falling in the M31 halo appears to survive longer than the other two fiducial cases: we attribute this behavior to the choice made for M31 of using the cloud mass found in \cite{afruni22}, which is slightly higher than the analytical estimate, as explained in more detail in Section~\ref{coolCloudsProps}. We therefore conclude that there does not seem to be a particular trend between the cloud survival and the mass of the halo where the cloud is infalling.

We have found that, irrespective of the mass of the halo in which they are accreting, within our mass range, cool CGM clouds infalling towards disc galaxies evaporate in the hot gas after an evaporation time ($t_{\rm{ev}}$) between 1.8 and 2 Gyr, at a distance from the disc ($z_{\rm{ev}}$) that is comprised within 0.5 and 0.6 $r_{\rm{vir}}$ (see Table~\ref{tab:resultsEvap}). This fundamental result, whose implications for galaxy evolution will be discussed extensively in Section~\ref{implications}, was obtained for our fiducial setup. In the following sections, we will examine how this result changes when varying the initial conditions of our simulations.

\subsection{Effect of thermal conduction on the cloud survival}\label{ResConduction}
In our fiducial simulations, we assumed that the suppression factor of the thermal conduction, which mimics the effect of a magnetic field in hampering the conduction efficiency, is equal to 0.1. This value is justified by the findings of previous works that modeled cool clouds moving in hot gas using MHD simulations \citep[see][]{kooij21}. However, it is not obvious that this value would hold also for the environment studied in this work, where we are simulating the external regions of a galactic halo, opposite to \cite{kooij21}, whose work is focused on a few kpc above the galactic disc.

   \begin{figure*}
   \includegraphics[clip, trim={0cm 0cm 0cm 0cm}, width=0.49\linewidth]{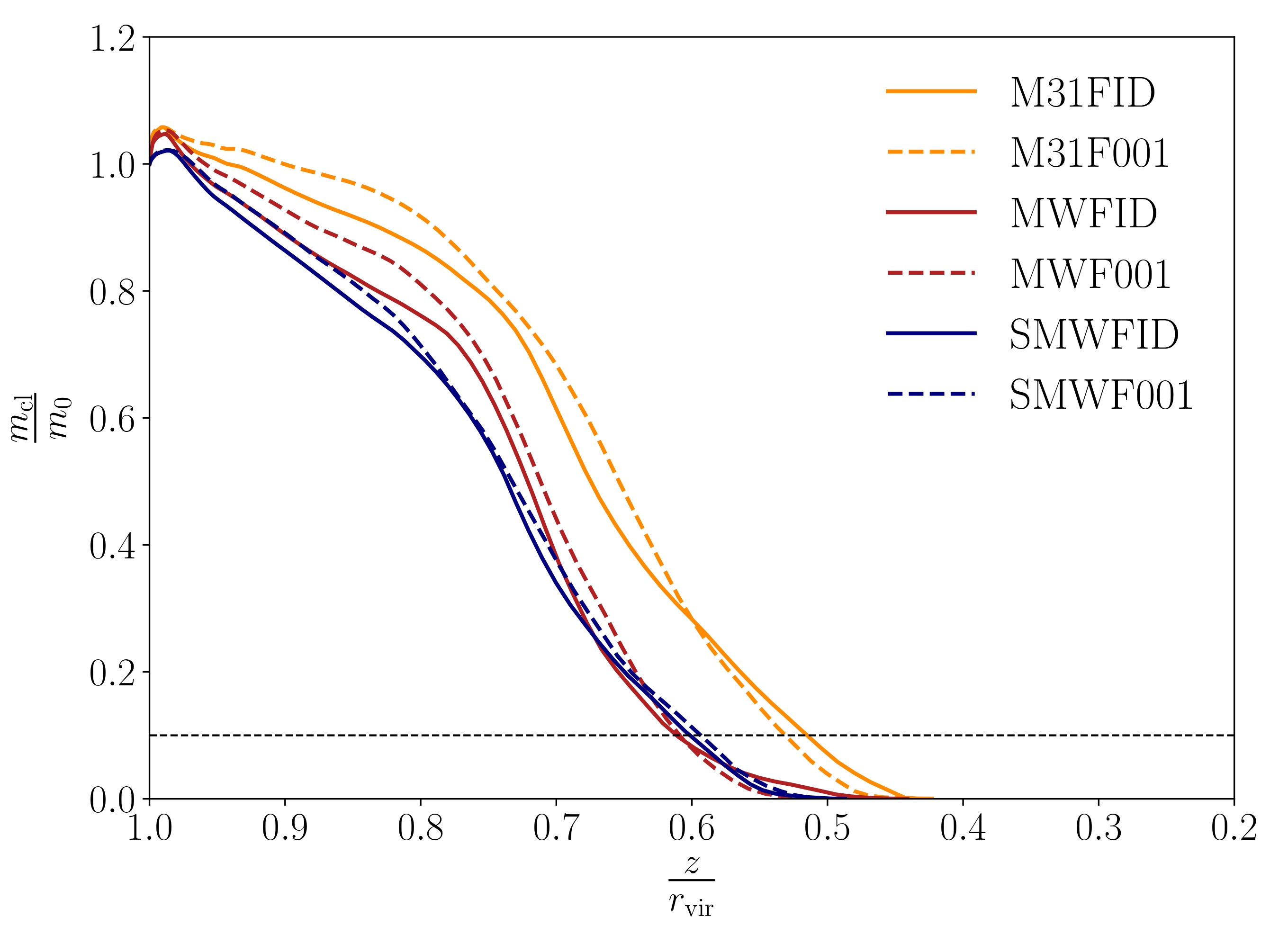}
   \includegraphics[clip, trim={0cm 0cm 0cm 0cm}, width=0.49\linewidth]{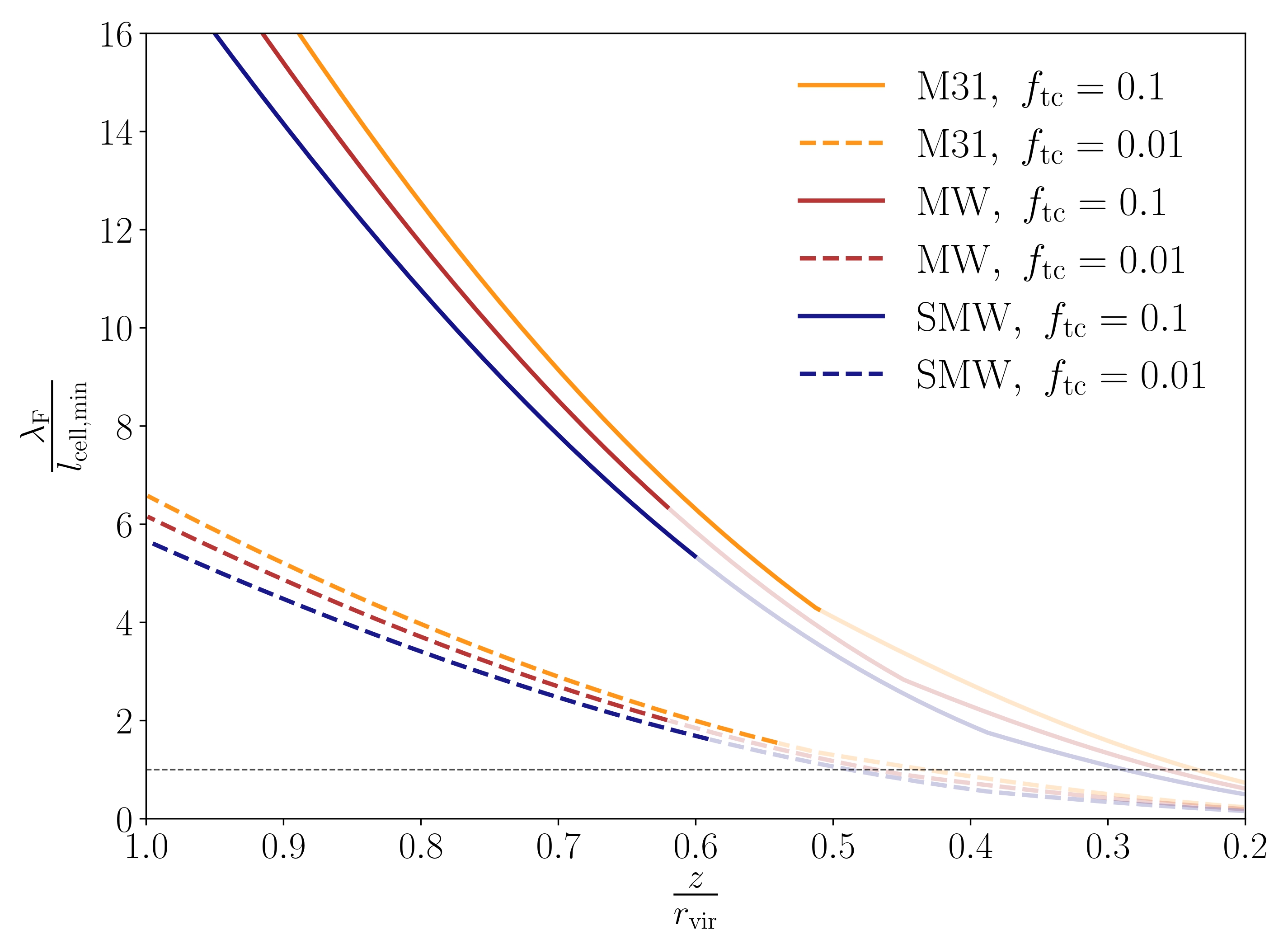}
   \caption{Left, evolution as a function of the normalized distance from the galaxy of the normalized cool gas mass present in the simulation box, for our fiducial simulations (solid curves) and for the simulations with a thermal conduction suppression factor $f_{\rm{tc}}=0.01$ (M31F001, MWF001, SMWF001, dashed lines). The dotted horizontal line is the same as in Figure~\ref{fig:profilesFID}. Right, values of the Field length, calculated through equation~\eqref{eq:lambdafield} and normalized to the smallest cell size of our fiducial simulations ($l_{\rm{cell,min}}=62.5$ pc), as a function of the normalized distance from the central galaxy, for the same simulations shown in the left panel. The dashed, horizontal black line marks $\lambda_{\rm{F}}=l_{\rm{cell,min}}$: the Field length is larger than $l_{\rm{cell,min}}$ everywhere in the region of the simulations where the cool gas is still present. All the curves have a higher level of transparency for $z<z_{\rm{ev}}$, where $z_{\rm{ev}}$ is the distance where the cloud has lost 90\% of its initial mass.}
              \label{fig:profilesf001}%
    \end{figure*}

To assess the influence of the choice of the suppression factor on our results, we ran three additional simulations, with the same initial conditions as the fiducial setup, except for a suppression factor $f_{\rm{tc}}=0.01$. This is 10 times smaller than the fiducial value and three times smaller than the lower limit allowed by the findings of \cite{kooij21}. We consider this case an extreme scenario, where the magnetic field is extremely efficient in reducing the conduction: we deem very unlikely the possibility that thermal conduction is in reality even less efficient than what simulated in these tests. The left-hand panel of Figure~\ref{fig:profilesf001} is similar to the bottom panel of Figure~\ref{fig:profilesFID}, showing the normalized cool gas mass as a function of the normalized distance from the disc for the fiducial cases and, in addition, also for the simulations with a reduced thermal conduction M31F001, MWF001 and SMWF001 (dashed curves).\footnote{The profiles as a function of time also show a very similar trend as in the fiducial simulations.} It is evident how the change in the value of the thermal conduction suppression factor does not affect significantly the results: the solid and dashed curves are almost overlapping with each other and, accordingly, the values of $t_{\rm{ev}}$ and $z_{\rm{ev}}$ are also very similar to the case with $f_{\rm tc} = 0.1$, as can be seen in Table~\ref{tab:resultsEvap}. Therefore, we can conclude that the survival of the cloud does not depend strongly on the suppression factor of thermal conduction. This might be due to the dual way in which thermal conduction affects our simulations (see also Section~\ref{intro}). A lower efficiency of thermal conduction, which decreases the efficiency of the heat exchange between the two gas phases and therefore should lead to a longer lifetime of the cool gas, also allows an easier development of the hydrodynamical instabilities that tend to destroy the cloud. The balance between these two opposite effects results in a very similar evolution of the cool gas in simulations independently on the exact value of $f_{\rm{tc}}$.

The reliability of the above finding can be assessed by considering the Field length \citep[][]{field65}:
\begin{ceqn}
\begin{equation}\label{eq:lambdafield}
\lambda_{\rm{F}}= \sqrt{\frac{f_{\rm{tc}}k_{\rm{Sp}}T_{\rm{cor}}}{n^2_{\rm{cool}}\Lambda_{\rm{net}}(T_{\rm{cool}}, Z_{\rm{cool}}, n_{\rm{cool}})}},
\end{equation}
\end{ceqn}
where $\Lambda_{\rm{net}}$ is the net cooling rate, $T_{\rm{cool}}$,  $Z_{\rm{cool}}$ and $n_{\rm{cool}}$ are the temperature, metallicity and density of the cool gas\footnote{In the following considerations, for the sake of simplicity, we assume that $T_{\rm{cool}}=2\times10^4$ K and $Z_{\rm{cool}} = 0.05 Z_{\odot}$ (as in the initial conditions), while we allow $n_{\rm{cool}}$ to vary with $z$, assuming pressure equilibrium with the initial hot corona density and temperature distributions (equations~\ref{eq:corn} and \ref{eq:corT}).}, $T_{\rm cor}$ is the temperature of the corona in K, and
\begin{ceqn}
\begin{equation}\label{eq:kspiChap5}
\kappa_{\rm{Sp}}=1.84\times10^{-5}\frac{T_{\rm{cor}}^{5/2}}{\ln \Psi}\ \rm{erg}\ \rm{s}^{-1}\ \rm{K}^{-1}\ \rm{cm}^{-1}\ ,
\end{equation}
\end{ceqn}\\
is the classical conductivity coefficient \citep{spitzer62}, where $\ln \Psi$ is the Coulomb logarithm ($\approx30$ in our setup). Following e.g.\ \cite{begelman90}, radiative processes dominate over thermal conduction if the scale of the considered gas structure is larger than the Field length, while on smaller scales thermal instability is suppressed and thermal conduction dominates. Based on this, any cool gas structures significantly smaller than the Field length should be rapidly destroyed by conduction and therefore we expect that $\lambda_{\rm{F}}$ sets a natural scale for the required resolution of our simulations \citep[see also, for example][]{armillotta16}.

In the right-hand panel of Figure \ref{fig:profilesf001} we show the dependence of the Field length on the distance from the disc, for the three halos considered in this study and for the two different choices of $f_{\rm{tc}}$. We can see that the Field length decreases with decreasing distance from the galactic disc, due to the increasing density of the cool cloud. Moreover, it is also clear how $\lambda_{\rm{F}}$ depends on the suppression factor: with lower values of $f_{\rm{tc}}$, the scales where thermal conduction dominates are also smaller, in proportion to $\sqrt{f_{\rm{tc}}}$. 
To assess the reliability of our results in terms of numerical resolution, we normalized $\lambda_{\rm{F}}$ with the length of the smallest cell size of our fiducial resolution setup (62.5 pc). We can see that, down to about half of the virial radius (after which there is no more cool gas in our simulation box), our fiducial resolution is sufficient to resolve the Field length and therefore properly capture the effect of thermal conduction, for both suppression factors. We will come back to resolution effects in Section \ref{convergence}.

\subsection{Varying the mass of the corona}\label{ResCoronalMass}
\begin{figure}
   \includegraphics[clip, trim={0cm 0cm 0cm 0cm}, width=\linewidth]{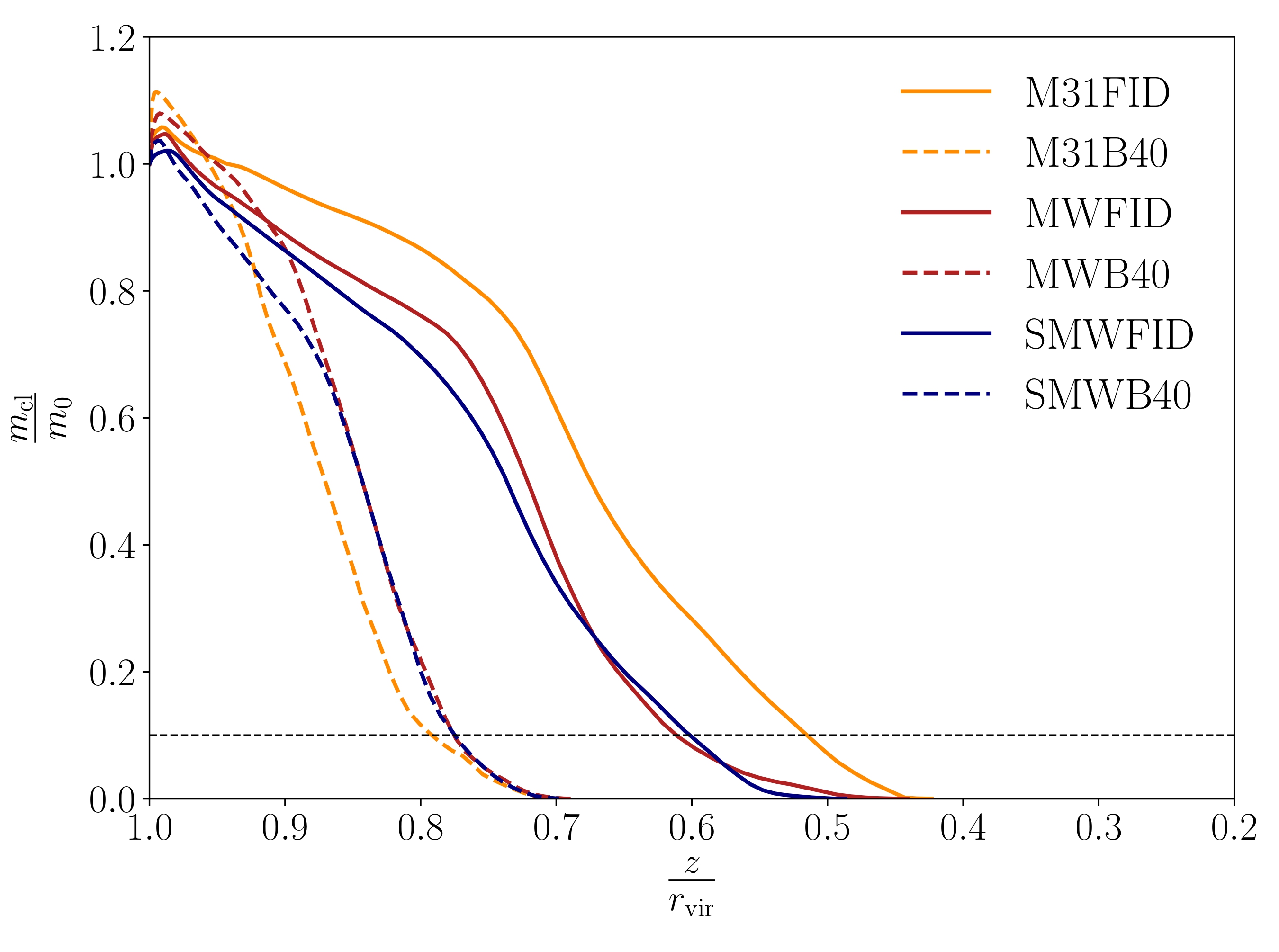}
   \caption{Similar the left panel of Figure~\ref{fig:profilesf001}, but here the dashed curves represent the simulations where the corona has twice the mass of our fiducial runs, $f_{\rm{cor}}=0.4$ (M31B40, MWB40 and SMWB40).}
              \label{fig:profiles2cor}%
    \end{figure}

In Figure~\ref{fig:profiles2cor}, we show the behavior of the cool gas mass with a different choice for the total mass of the hot corona. The solid curves represent again the fiducial simulations as in Figure~\ref{fig:profilesFID}, while the dashed curves show the results of the simulations M31B40, MWB40 and SMWB40, where we assumed that the hot CGM has a mass equal to 40$\%$ of the total baryonic mass within the galaxy halo, twice the amount present in our fiducial simulations. Recent works \citep[e.g.][]{nicastro23} have suggested that the majority of the missing baryons \citep[][]{mcgaugh} may reside in the hot phase, therefore it is possible that with our fiducial choice we are underestimating the amount of hot medium present in the halo. This may impact our results in two ways. On the one hand, higher densities lead to faster cooling and could potentially lead to a longer survival time for the clouds; on the other hand, we have seen in Section~\ref{coolCloudsProps} that a more massive corona implies more numerous and less massive clouds in order to be consistent with observational constraints on the cool gas covering factor (see Section \ref{coolCloudsProps} and Table~\ref{tab:inCond})\footnote{The intuitive explanation for this is that a more massive corona exerts a higher pressure and therefore compresses clouds of a given mass to a smaller size, leading to a lowered covering factor. This effect can be compensated by assuming less massive (and more numerous) cloudlets.}, and less massive clouds are potentially easier to destroy. It is therefore important to explore this scenario to assess which of the two effects mentioned above prevails over the other.

We can see from Figure~\ref{fig:profiles2cor} that clouds embedded in a more massive corona evaporate even more rapidly than in our fiducial case and have lost the majority of their mass at $\sim0.8r_{\rm{vir}}$ (see also Table~\ref{tab:resultsEvap}). We therefore see that the effect of a smaller cloud mass prevails, leading, for all the considered halo masses, to a more rapid evaporation compared to the fiducial case (see also Table~\ref{tab:inCond}). These experiments indicate that, even if the corona is more massive than what assumed in our fiducial simulations, cool clouds whose properties are in agreement with the current observational constraints (i.e.\ the covering fraction of the cool CGM), are going to evaporate into the hot CGM at very large distances from the galactic disc.

\subsection{Varying the cloud velocity}\label{resVelocity}
   \begin{figure*}
   \includegraphics[clip, trim={0cm 0cm 0cm 0cm}, width=0.49\linewidth]{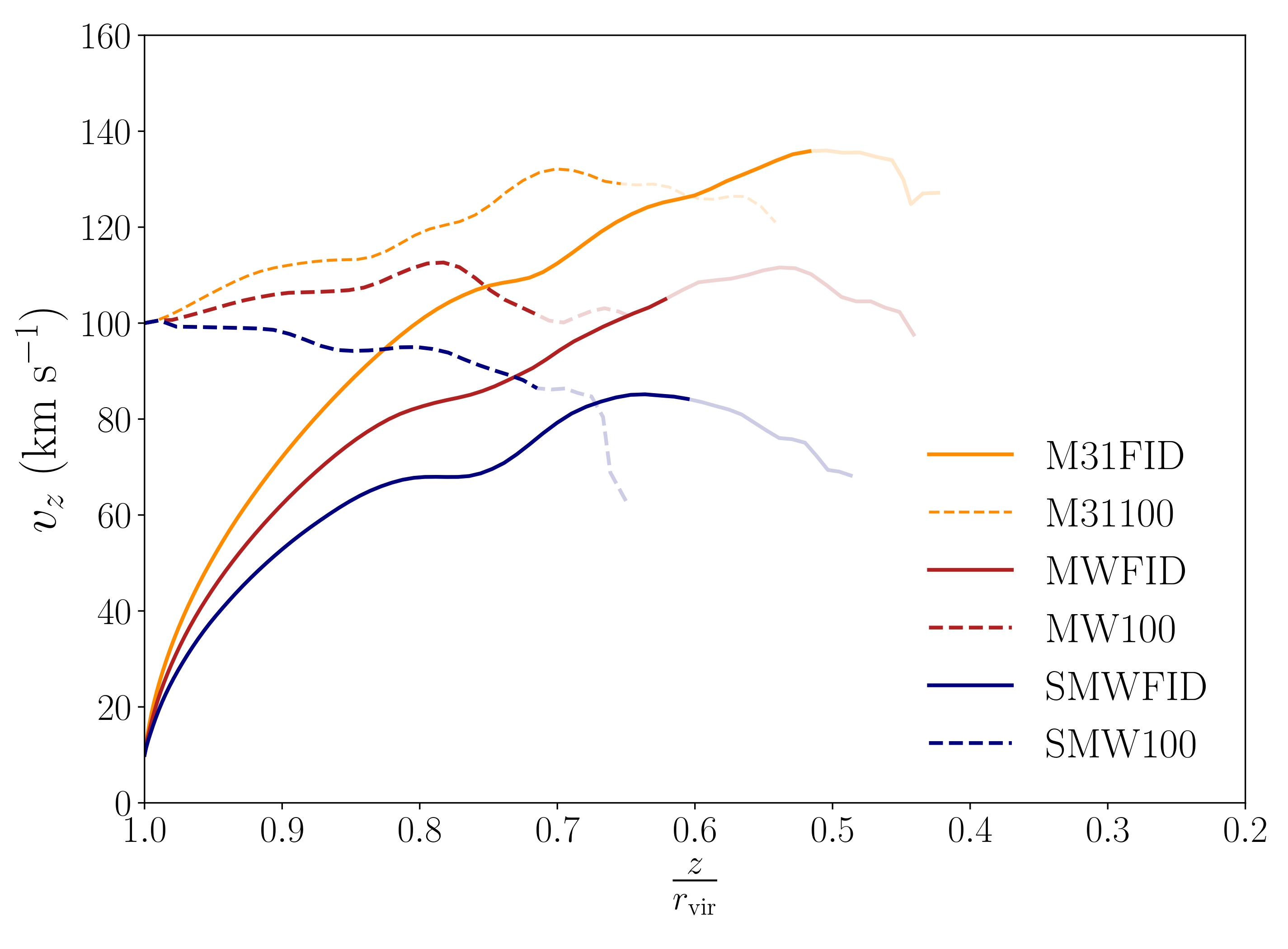}
   \includegraphics[clip, trim={0cm 0cm 0cm 0cm}, width=0.49\linewidth]{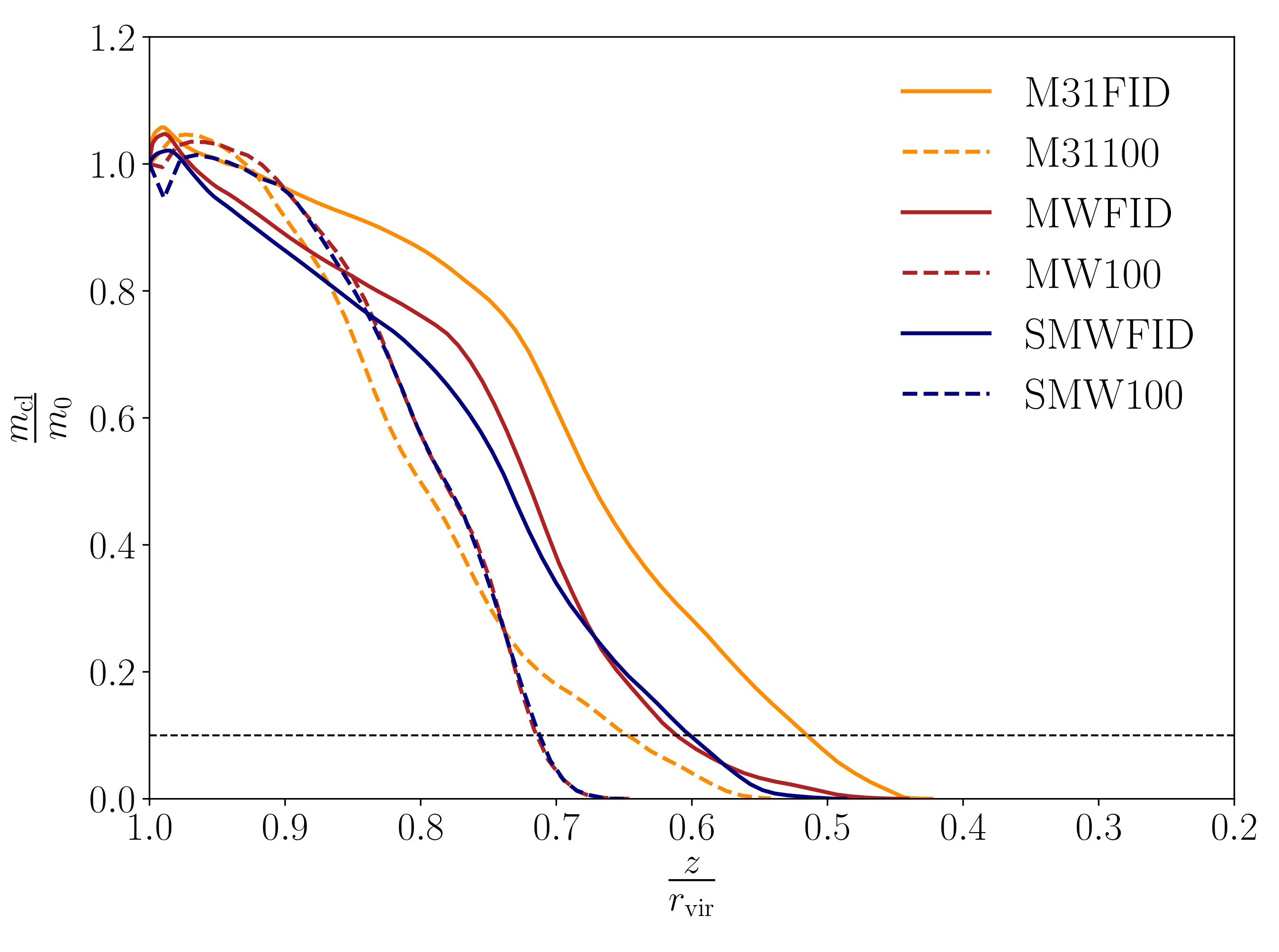}
   \caption{Left, mass-weighted average velocity of the cool gas present in the simulation box, as a function of the normalized distance from the galactic disc, for our fiducial simulations (solid curves) and for the simulations M31100, MW100 and SMW100 (dashed curves). All the curves have a higher level of transparency for $z<z_{\rm{ev}}$, where $z_{\rm{ev}}$ is the distance where the cloud has lost 90\% of its initial mass. Right, analogous to Figure~\ref{fig:profiles2cor}, but for the same simulations as in the left-hand panel.}
              \label{fig:profilesvel}%
    \end{figure*}
For our fiducial scenario, we assumed that the clouds are originated by the destruction of the infalling filaments when they encounter the pre-existing hot corona. This process would significantly slow down the infalling gas, resulting in a low initial velocity for the cool clouds. We therefore chose an infall velocity of $10$ km s$^{-1}$, which is also justified by the results of the analytical study of \cite{afruni22}, focused on the halo of M31 and interestingly in agreement with the results of \cite{afruni19} on a sample of early-type galaxies. To assess the role of this assumption, we have also run some simulations where the cloud initial velocity is higher than in our fiducial case. In this modified scenario, the filament fragmentation is not accompanied by a strong deceleration of the cool gas. In particular, we ran simulations where the cloud initial velocity is 100 km s$^{-1}$, 10 times higher than the fiducial case and closer to the virial velocity $v_{\rm{vir}}= \sqrt{GM_{\rm{vir}}/r_{\rm{vir}}}$ of our three halos, which we estimated being equal to 120, 140 and 160 km s$^{-1}$, respectively for the Sub-MW, MW and M31 halos. This is also in agreement with some cosmological simulations, which find that accreting cold filaments typically reach infall velocities equal to a fraction (between 0.6 and 0.8, depending on the halo mass and redshift) of the virial velocity \citep[see][]{goerdt15}. Our adopted value of 100 km s$^{-1}$ is therefore compatible with the expected filament infall velocity in the absence of significant decelerations from unresolved hydrodynamical processes. Note again that, based on the considerations of Section~\ref{coolCloudsProps}, the initial masses of the clouds in these simulations are lower with respect to the respective fiducial cases (see also Table~\ref{tab:inCond}), in order to remain consistent with the constraints of the observed cool gas covering factor around star-forming galaxies.\footnote{In this case, the intuitive explanation is that, were the cloud mass kept fixed, a higher velocity would imply a lowered probability for each cloud to be intercepted at any given time along a given line of sight, leading to a reduction in the observed covering factor.}

In Figure~\ref{fig:profilesvel} we can see the results of the high-velocity simulations (dashed curves), compared to the fiducial cases (solid curves). On the left-hand panel we show the cool gas velocity as a function of the normalized distance from the galactic disc. We marked with a higher level of transparency the region where the clouds have already lost 90$\%$ of their mass ($z<z_{\rm{ev}}$). We recall that after this point the cloud is entirely fragmented by the interactions with the corona and almost completely evaporated in its hot surroundings. We can see that in the fiducial simulations the clouds are initially strongly accelerated by the gravitational force, while later on the acceleration is significantly reduced, as a consequence of the drag force, which opposes gravity, and the onset of hydrodynamical instabilities. The behavior of the clouds starting with a high velocity is instead different: the clouds are never significantly accelerated during their infall, as the drag force tends, since the beginnning, to counter-balance the gravitational attraction. In all cases, the cool gas never reaches infall velocities higher than about 140 km s$^{-1}$. Note that the infall motion is always subsonic, as the hot gas sound speed at the virial radius varies from about 110 to almost 150 km s$^{-1}$ for the three considered halo masses, and increases at lower galactocentric radii due to the higher temperatures (see Figure \ref{fig:Hotgas}).

In the right-hand panel of Figure~\ref{fig:profilesvel} we show instead the evolution of the cool gas mass within the simulation box. It is clear that also in the case of clouds starting with a higher initial velocity, the fate of accreting cool CGM clouds is to evaporate into the hot corona. The rate of evaporation is even faster than the fiducial case, with $t_{\rm{ev}}$ being shorter than 1 Gyr and $z_{\rm{ev}}$ larger than 0.6 $r_{\rm{vir}}$ (Table~\ref{tab:resultsEvap}). We therefore conclude that our main result, i.e.\ the evaporation of infalling cool clouds at large distances from the disc, holds also when we consider a higher initial velocity of the IGM clouds.
{ 
 \begin{table}
\begin{center}
\begin{tabular}{*{6}{c}}
(1)&(2)&(3)&(4)&(5)&(6)\\
\hline  
\hline
 Sim. Id & $t_{\rm{ev}}$ & $z_{\rm{ev}}$ & $z_{\rm{ev}}/r_{\rm{vir}}$ & $t_{\rm{ev}}$ & $z_{\rm{ev}}$ \\
     & (Gyr) & (kpc)  &  & (Gyr) & (kpc) \\ \\
 & & & & $T<10^5$ K & $T<10^5$ K \\
\hline 
M31FID & $2.01$ & $174$ & $0.52$ & $2.07$& $166$\\
M31F001 &$1.95$ & $180$& $0.54$ & $2.01$& $173$\\
M31B40 & $1.32$ & $266$& $0.79$ & $1.41$& $259$\\
M31100 & $0.99$ & $220$& $0.65$ & $1.08$& $208$\\
MWFID & $1.77$ & $183$& $0.62$ & $1.89$& $167$\\
MWHRES & $1.74$ & $186$ & $0.63$ & $1.80$& $180$\\
MWLRES & $2.04$ & $152$ & $0.52$ & $2.16$& $138$\\
MWF001 & $1.77$ & $182$& $0.62$ & $1.86$ & $173$\\
MWB40 & $1.35$ & $230$& $0.78$ & $1.44$& $224$\\
MW100 & $0.78$ & $211$& $0.72$ & $0.87$& $202$\\
SMWFID & $1.83$ & $152$& $0.60$ & $1.95$& $143$\\
SMWF001 & $1.83$ & $151$ & $0.60$ & $1.95$& $141$\\
SMWB40 & $1.38$ & $196$ & $0.78$ & $1.50$& $189$\\
SMW100 & $0.75$ & $180$ & $0.71$ & $0.84$& $173$\\
\hline
\end{tabular}
\end{center}
\captionsetup{}
\caption[]{Summary of the main results of the simulations performed in this work. Columns (2), (3) and (4) show respectively the values of $t_{\rm{ev}}, z_{\rm{ev}}$ and $z_{\rm{ev}}/r_{\rm{vir}}$, indicating the time and distance from the central galaxy after which the cool gas with $T<3\times10^4$ K has effectively evaporated into the corona, while columns (5) and (6) show the values of $t_{\rm{ev}}$ and $z_{\rm{ev}}$ for the gas with $T<10^5$ K.}\label{tab:resultsEvap}
   \end{table}
   }
\subsection{Convergence study}\label{convergence}
  \begin{figure*}
   \includegraphics[clip, trim={0cm 0cm 0cm 0cm}, width=\linewidth]{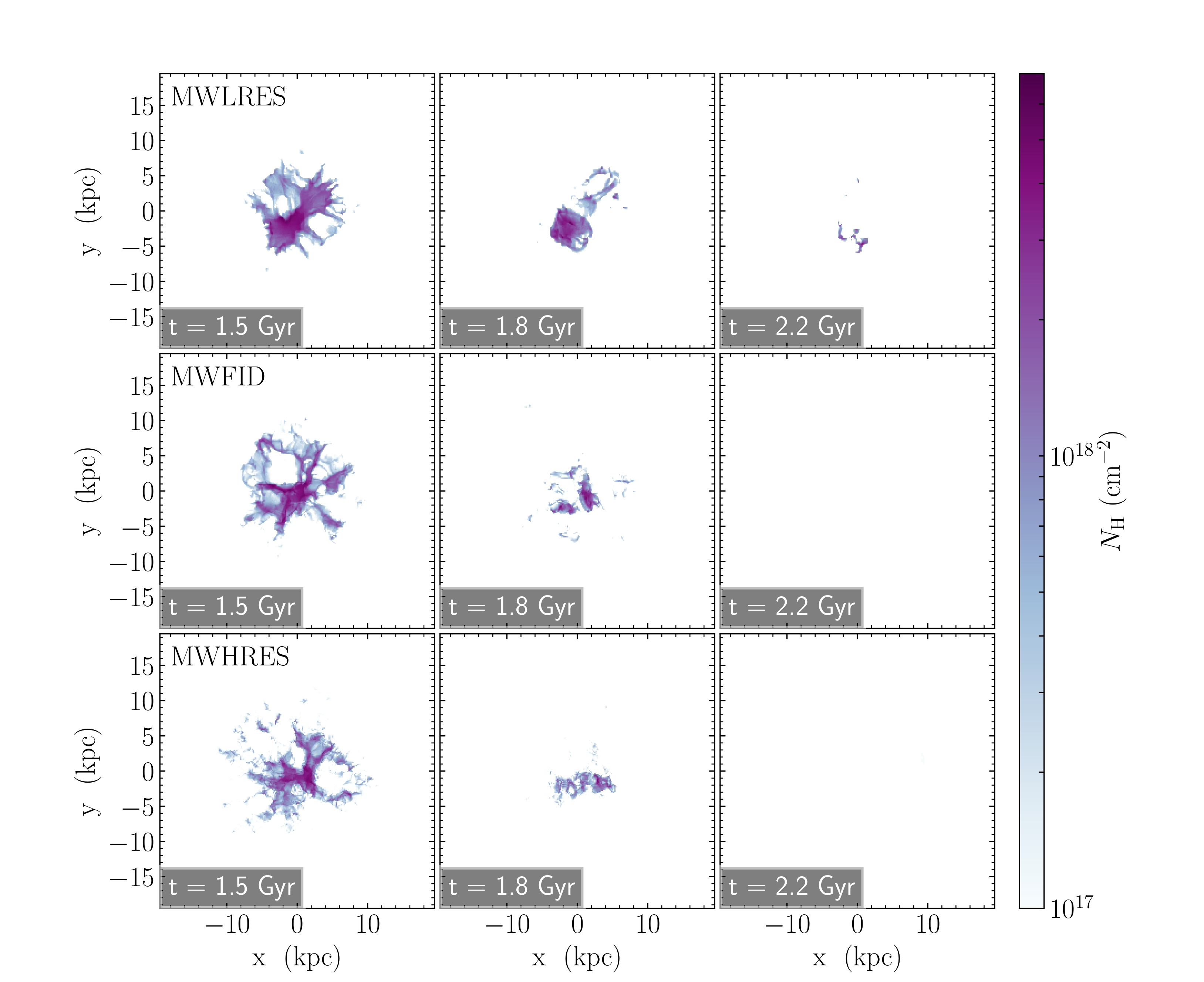}
   \caption{Total hydrogen column density, integrated along the $z$-axis, of the cool gas ($T<3\times10^4$ K), in a zoom-in of the simulation box that contains all the cool gas, at three different times (1.5, 1.8 and 2.2 Gyr), for the simulations MWLRES (top panels), MWFID (middle panels) and MWHRES (bottom panels). Note how MWFID and MWHRES show a very similar cool gas evolution, while the cloud evaporates more slowly in MWLRES.}
              \label{fig:SnapshotResolutions}%
    \end{figure*}
We have seen in Section~\ref{ResConduction} that our fiducial simulations have a maximum resolution that is smaller than the Field length, which implies that our results should be converged in terms of total amount of cool gas. To further test the convergence of our numerical experiments, we ran two additional simulations, with the same initial conditions as our fiducial simulation for the MW halo, MWFID, but refining one level more (MWHRES, with the highest resolution being 31.25 pc) and one less (MWLRES, with highest resolution equal to 125 pc).

In Figure~\ref{fig:SnapshotResolutions}, we compare the morphology of the simulated clouds at the three explored resolution levels, as a function of time. For a better view of the structures in which the cloud is fragmented, we use a projection `from below': in particular,  we show the total hydrogen column density of the cool gas ($T<3\times10^4$ K), projected along the $z$-direction, at three times in the cloud evolution (1.5, 1.8 and 2.2 Gyr) and for the simulations at the three different resolutions mentioned above. We can note how the cloud evolution is qualitatively the same for the three cases and that in the two simulations with the highest resolution (MWFID and MWHRES) the rate of evaporation of the cloud is very similar, with the cool gas completely evaporated in the hot corona for $t = 2.2$ Gyr. This confirms that our fiducial simulations have already reached convergence. Increasing the resolution increases (as expected) the number of small substructures, but given that these are smaller than the Field length they almost immediately evaporate into the hot gas and do not significantly change the final result. In the lowest-resolution simulation (MWLRES), instead, the cloud is fragmented into larger chuncks, which tend to evaporate more slowly, so that for $t= 2.2$ Gyr some cool material is still present in the simulation box. This shows that lowering too much the resolution has the effect of artificially increasing the survival time of the cool cloud. It is important to keep this in mind when considering the cool gas evolution in cosmological simulations. We will come back to this point in more detail in Section~\ref{disc:comparison}.

\begin{figure}
   \includegraphics[clip, trim={0cm 0cm 0cm 0cm}, width=\linewidth]{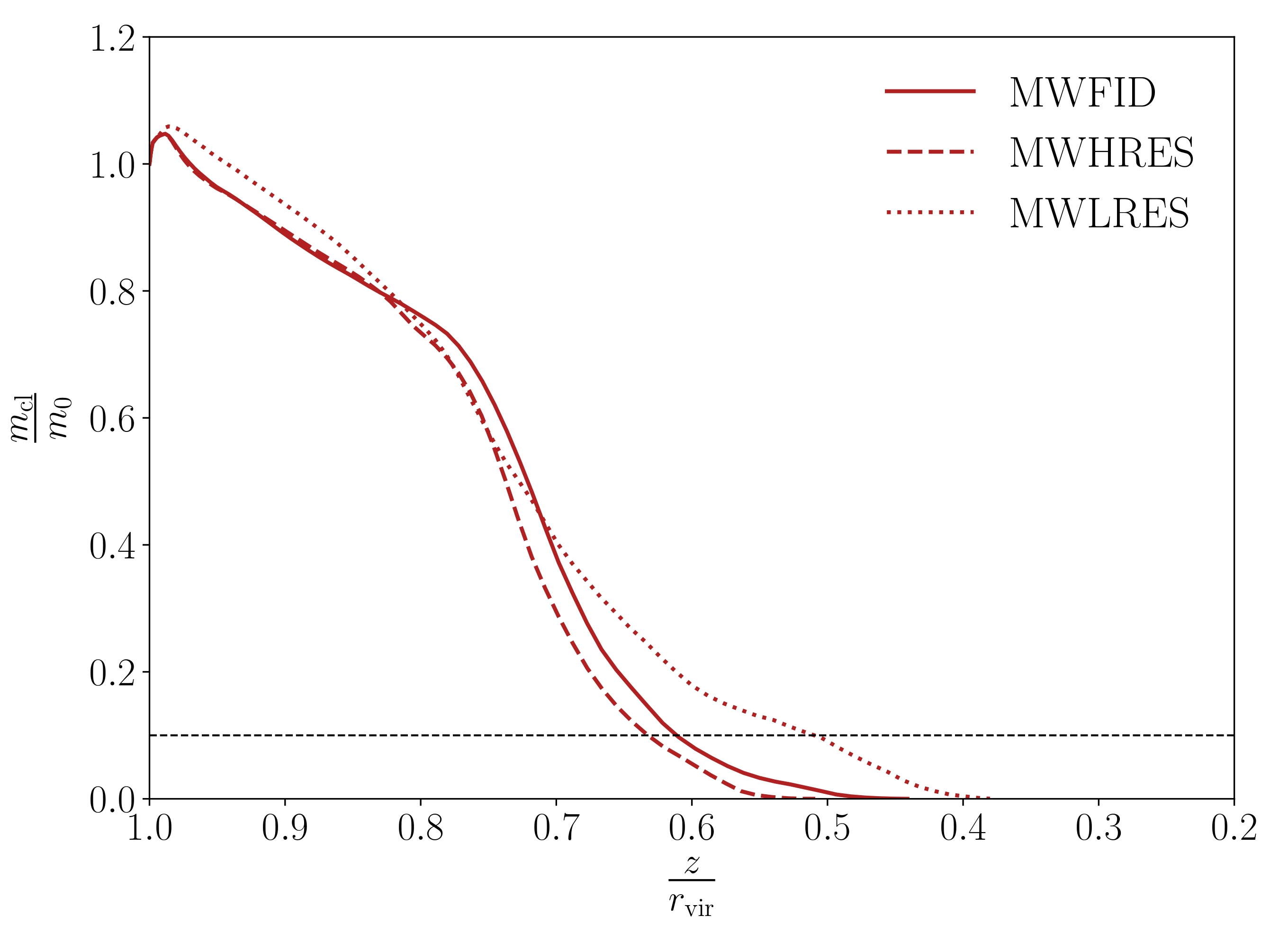}
   \caption{Evolution with respect to the normalized distance from the central galaxy of the cool ($T<3\times10^4$ K) gas mass present in the simulation box, normalized to the initial cloud mass, for the three simulations of a MW-like halo at fiducial (MWFID), higher (MWHRES) and lower (MWLRES) resolution. The horizontal dotted line marks the level where the cloud has lost 90\% of its initial mass, i.\ e.\ $m_{\rm{cl}}/m_0=0.1$. We observe that a numerical resolution worse than 62 pc leads to an artificial increase of the survival time of the infalling cool gas.}
            \label{fig:profilesResolutions}%
    \end{figure}

More quantitatively, in Figure~\ref{fig:profilesResolutions} we show, for the same three simulations, the evolution of the cool gas mass as a function of the normalized distance from the galactic disc. In all cases the fate of the cloud is to completely evaporate into the hot corona, with the cool gas surviving longer in lower resolution simulations, in agreement with the results already shown in Figure~\ref{fig:SnapshotResolutions}. The two profiles of MWFID and MWHRES almost overlap with each other, while there is a delay in the cloud evaporation in the MWLRES simulation. This is also evident from the results reported in Table~\ref{tab:resultsEvap}, where we can see that both $t_{\rm{ev}}$ and $z_{\rm{ev}}$ are very similar for the two simulations with the highest resolutions, while they are respectively larger and smaller for the lowest resolution case. This analysis confirms that we are very close to convergence and we do not expect strong differences at even higher resolutions. Moreover, the small trend in Figure~\ref{fig:profilesResolutions} would suggest that, if any difference is present at higher resolution, the cloud would evaporate even more rapidly into the hot corona, only reinforcing the main result of this paper.

Finally, we may use our results to give an estimate of the number of cells by which the Field length needs to be resolved to achieve converged results in the presence of thermal conduction, at least for a set-up comparable to that of our simulations. In particular, we note that for $z\sim z_{\rm{ev}}$ (where the cloud is evaporating in the hot medium and $\lambda_{\rm{F}}$ reaches its minimum meaningful value), the Field length is still resolved by about 5 to 7 cells for the fiducial (reasonably converged) simulation MWFID, while the number of cells per Field length is only 2 or 3 in the (slightly discrepant) MWLRES. Based on this, we suggest that resolving the Field length with at least 5 to 7 grid cells seems advisable in order to ensure a good convergence in terms of cool gas mass evolution, at least for a problem similar to ours.\footnote{Note that, if this criterion is valid, then our simulations with reduced thermal conduction ($f_{\rm tc} = 0.01$) may not be perfectly converged yet (Figure \ref{fig:profilesf001}, right-hand panel).}

\section{Discussion}\label{discussion}
\subsection{Limitations of this study}\label{limitations}
The results of this work are based on our numerical simulations, which are affected by limitations, some of which have already been discussed throughout the previous sections.

One of the limitations of this study (and more in general of studies of cool clouds in the CGM) is an inherent difficulty in the choice of the initial conditions. In our case, the most critical factor is probably the absence of clear observational measurements of the mass of typical cool clouds in the proximity of the virial radius of star-forming galaxies. To overcome this limitation, we have resorted in Section \ref{coolCloudsProps} to an indirect estimate of plausible typical cloud masses, based on observations of the covering factor of cool CGM absorbers (see references in Section \ref{intro}). The reason why typical cloud masses and the covering factor are linked to each other is primarily geometrical and, in its essence, is the same reason why, for any given amount of water, a fog of many small droplets has a higher covering factor compared to a collection of a few more massive drops. The covering factor is itself somewhat uncertain. For this reason, we have resorted to the drastically conservative choice to identify the covering factor with the detection fraction of cool absorbers. As also discussed in Section \ref{coolCloudsProps}, this represents a lower limit to the true covering factor, as it ignores the possibility that multiple clouds lie along the same line of sight, also implying that our adopted cloud masses should be better understood as upper limits to the true typical cloud masses. To be more quantitative, we took advantage of the fact that, for some of the samples mentioned in Section \ref{coolCloudsProps} \citep[in particular][]{borthakur15, lehner20p}, we do have constraints on the number of kinematically distinct components along every line of sight. Using this information and following a procedure similar to \cite{afruni20,afruni22}, we have obtained improved vales for the covering factor $f_{\rm c} = 1.1\pm0.4$, $1.7\pm0.3$ and $1.0\pm0.5$ for respectively the M31, MW and Sub-MW halo. As expected, these are in general higher than the values given in Section \ref{coolCloudsProps}, but still consistent within Poissonian uncertainties, with the only exception of the MW halo, for which the difference is highly significant. In this case, a revised cloud mass would be $\log(m_{\rm{cl}}/M_{\odot})=5.7\pm0.2$, about 6 times smaller than the one adopted as our initial condition. Although we did not run simulations with this lower mass, it should be expected \citep[and is also confirmed by previous studies, e.g.][]{armillotta17} that less massive clouds are easier to disrupt compared to more massive ones. We therefore conclude that uncertainties in covering factor and typical cloud masses go in the direction of strengthening the result of the paper.

A second limitation is that all our results are based on the definition of cool gas as the medium with $T<3\times10^4$ K. While this assumption is justified by the typical temperatures estimated for the cool ionized CGM, this threshold is relatively arbitrary and our results could change if we considered also gas with slightly higher temperatures. In order to test this, we repeated our calculations adopting a different threshold of $T<10^5$ K, which is a value often used in the literature to define the cool CGM \citep[e. g.][]{armillotta17}. In the last two columns of Table~\ref{tab:resultsEvap} we report the values of $t_{\rm{ev}}$ and $z_{\rm{ev}}$ that we found using this new threshold for the temperature: these are, as expected, respectively slightly higher and lower than the fiducial values, but they never differ from the latter by more than 10\%. We conclude that our main result regarding the evaporation of the CGM clouds is not affected by the precise temperature threshold chosen for the definition of the cool gas.

Another slightly arbitrary choice that we have made is the assumption of a constant metallicity $Z=0.3\ Z_\odot$ throughout the hot corona. Our chosen value is justified based on estimates from X-ray observations \citep[e.g.][]{miller15, bregman18}. However, X-ray data generally probe only the corona at distances $\lesssim 50\ {\rm kpc}$ from the disc, while the metallicity at the distances considered in our simulations is, to date, unknown. It is possible that at large distances the hot gas may not be significantly enriched (for instance because of a lower impact of feedback from the central galaxy) and therefore its metallicity could be lower than our adopted value. A lower metallicity in the outer regions would however result in a lower cooling rate that reduces the survival time and would therefore go in the direction to strengthen our main finding of cloud evaporation.

One important simplification that we have made in this study is that, as already mentioned in Section~\ref{hotmedium}, we have assumed for simplicity that the hot gas does not rotate. This is in contrast with the established notion that the corona must have significant angular momentum, as supported by both theoretical arguments \citep[e.g.][]{teklu15, pezzulli17,stern19,afruni22} and observational evidence in the Milky Way \citep[][]{hodges16}. The approximation appears nonetheless justifiable if one considers that, at the large distances of interest for our study, a significant angular momentum can be naturally achieved with modest rotational velocities. For instance, the model of the corona of M31 presented in \citet{afruni22} predicts that, at a galactocentric distance of $\sim$ 170 kpc (where the cool cloud evaporates in our fiducial model applicable to this galaxy, see Table \ref{tab:resultsEvap}) the rotation velocity is about 20 km s$^{-1}$, i.e.\ just a small fraction of the infalling terminal velocity of the evaporating cloud ($\sim$ 120 km s$^{-1}$, see Figure \ref{fig:profilesvel}). We therefore expect that the simplification of a non-rotating corona should not significantly affect our results, although further tests on this topic may be an interesting experiment for future work.

We have further neglected the effect of cloud self-gravity, focusing instead on the external gravitational field. This approximation seems justified in the light of the existing literature. \cite{li20SG} found that self-gravity has a negligible effect, compared to thermal conduction or radiative cooling, for clouds with masses lower than the Jeans mass. This condition is fully satisfied for the parameters of our simulations, for which the Jeans mass, for any $z > 0.5 r_{\rm vir}$, is higher than $10^{10}\ {\rm M}_{\odot}$, i.e.\ more than three orders of magnitude higher than our cloud masses. On the other hand, \cite{sander21} have pointed out that self-gravity may be potentially relevant in influencing the cloud lifetime by reducing the development of RT instability and non-negligible effects have been reported for clouds with masses as low as 10\% of the Jeans mass \citep{sander2019}. However, even 10\% of the Jeans mass is still more than two orders of magnitude larger than our clouds. Therefore we are confident that including self-gravity would only have a minor impact on our results.

As also already mentioned, we have neglected the detailed effects of the magnetic field \citep[see for example][]{li20SG,liang20,sparre20}, apart from the suppression of thermal conduction. This choice was made because including magnetic fields would have increased the computational cost of our experiments and also because we have, to date, no observational constraints on the magnetic field in the outer regions of the halos of disc galaxies. Including a coronal magnetic field with a strength that varies from 0.1 to 1 $\mu$G would tend to reduce the stripping and the mixing of hot and cool gas \citep[e.g.][]{gronnow18}. However, this effect would presumably be small in the outskirts of the corona, where the field is expected to be weak, of the order of 0.01 $\mu$G \citep[see for example][]{marinacci18}. Again following \cite{gronnow18}, such a weak field would not significantly affect the evolution of the cool gas.

Finally, in our simulations the clouds are initially spherical, which is an approximation that is probably far from the real configuration of cool clouds at the virial radius of galaxies. In particular, we assumed here that the presence of such clouds is due to the fragmentation of the infalling filaments during the interaction with the pre-existing hot corona, where the IGM decouples from the infalling dark matter due to hydrodynamical effects. The initial shape of the clouds will strongly depend on the details of this process, which are to date not well understood. Characterizing the filament fragmentation can be challenging for cosmological simulations, because of resolution limitations (as discussed better in Section \ref{disc:comparison}). High-resolution simulations of inflowing filaments \citep[e.g.][]{mandelker20} can be useful in this sense, but are necessarily idealized and cannot easily be connected to a cosmological context. Developing ad-hoc simulations that have the necessary resolution to follow the fragmentation of cosmological filaments, while outside the scope of this work, is a promising direction for future investigation.

\subsection{Comparison with previous studies}\label{disc:comparison}
In this paper, we have investigated whether cool clouds accreting from the IGM can survive their journey and feed the star formation of the central galaxy before evaporating into the hot corona. Previous studies have tried to answer very similar questions.

In particular, \cite{nipoti07} investigated this problem from an analytical point of view, focusing on the effect of thermal conduction on the evaporation of cool clouds accreting from the IGM through a hot corona. More in detail, they compared the conduction time, which for a spherical cloud can be defined as \citep[see also][]{spitzer62,cowie77}
\begin{ceqn}
\begin{equation}\label{eq:tcond}
t_{\rm{cond}}= \frac{25k_{\rm{B}}n_{\rm{cool}}r_{\rm{cl}}^2}{12f_{\rm{tc}}k_{\rm{sp}}}\ ,
\end{equation}
\end{ceqn}
with the dynamical time that is necessary for the cloud to reach the galaxy. They found that conduction is very efficient in making cool clouds evaporate in the halos of massive elliptical galaxies, providing also a possible explanation for the absence of star formation therein \citep[see also][]{afruni19}. The accretion of cool clouds should instead happen more easily in lower mass galaxies, where star formation is more often still ongoing. We used this analytical framework to calculate the conduction time for the initial setup of our fiducial simulations. We obtained that, at the virial radius, $t_{\rm{cond}}\sim30, 60$ and $120$ Gyr, respectively for M31FID, MWFID and SMWFID. These timescales are therefore much larger than the dynamical time, which is of a few Gyr. According to this estimate, one would conclude that clouds are able to accrete onto the central galaxy, while we have found in our simulations that the cool gas evaporates into the hot corona in about $2$ Gyr or less, more than 100 kpc away from the disc. The reason for this discrepancy is due to the fact that our simulations also take into account the development of hydrodynamic instabilities. In particular, our simulations show that hydrodynamic effects can lead, in less than one dynamical time, to the fragmentation of the cloud down to scales as small as 200 pc, for which the conduction time is as short as 50 Myr, drastically reducing the ability of a cloud to survive evaporation compared to the case in which the clouds remains monolithic and spherical. We refer to Section \ref{subsec:processes} below for a discussion of the main hydrodynamic processes involved and their associated timescales.

Also in contexts other than IGM accretion, several studies have analyzed the evolution of cool clouds moving through a hot medium using high-resolution simulations similar to those presented in this paper. Many of these works have found that, in the presence of radiative cooling, cool clouds can acquire (rather than lose) mass, through the cooling and condensation of the gas at intermediate temperatures that forms from the mixing of coronal gas with layers of the clouds stripped as a consequence of hydrodynamic instabilities \citep[e.g.][]{marinacci10accr, armillotta16, gronke18,gronke20,kooij21,tan23}. We believe that our findings are not in contradiction with these results and that the difference is due to the fact that the studies mentioned above were focused on relatively small distances (within about 10 kpc) from the galactic disc, where the properties of the cool and hot media are different than those probed by our simulations. \cite{armillotta17} have also studied the evolution of clouds at galactocentric distances of $\sim 50-150$ kpc, using 2D simulations at high-resolution (2 pc) and including the effect of cooling, an ionizing UV background and isotropic thermal conduction. It is not easy to compare our results to those of \cite{armillotta17}, because their set-up did not include a gravitational field and (perhaps most importantly) because they followed the evolution of their clouds for only 250 Myr (compared to the 3 Gyr of our simulations). However, we can notice that, limited to this short time span, all our clouds survive the interaction with the corona, in agreement with the simulations of \cite{armillotta17} with sufficiently large cloud mass ($M_{\rm cl} \geq 2 \times 10^4 \ {\rm M}_\odot$, which also applies to all our simulations). 

\begin{figure}
   \includegraphics[clip, trim={0cm 0cm 0cm 0cm}, width=\linewidth]{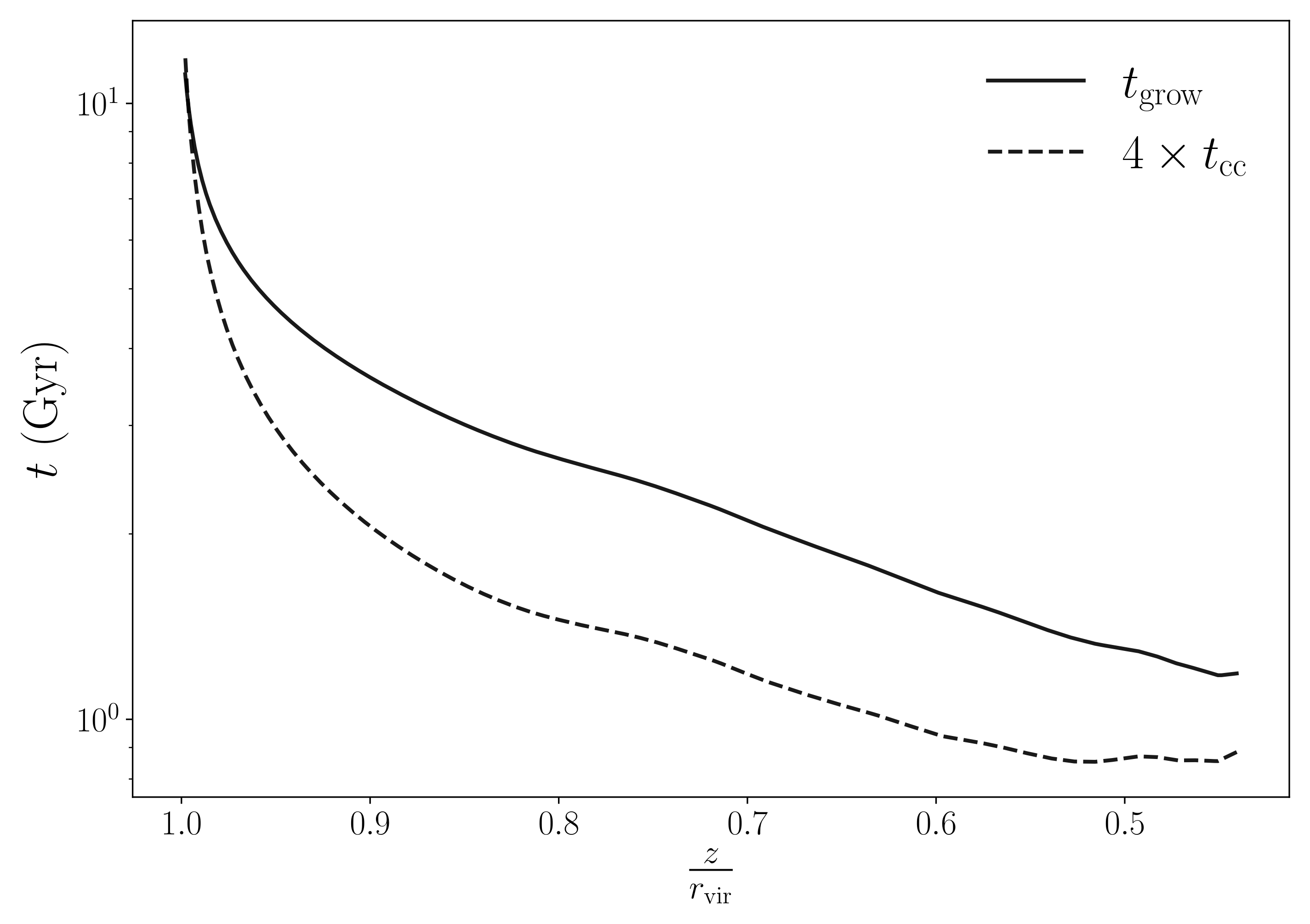}
   \caption{Timescales of growth and destruction of the cool cloud, as a function of the normalized distance from the galactic disc, calculated using the prescriptions by \protect\cite{tan23}. Cool clouds in our simulations fulfill the criterion $t_{\rm{grow}} > 4 t_{\rm{cc}}$, in agreement with the fact that they are not accreting cool gas mass from the hot corona.}
              \label{fig:tgrowtcc}%
    \end{figure}

We can look for a quantitative criterion to capture the very different behaviour of clouds in the outer and inner regions of the CGM of galaxies. A relevant timescale to consider is the time that is needed for the cloud to be destroyed by hydrodynamical effects and its comparison with the time necessary for the cloud to grow. To this purpose, \cite{gronke18} introduced a criterion to determine whether a cool cloud entrained in hot gas is expected to grow, defined as:
\begin{ceqn}
\begin{equation}\label{eq:tcool_tcc}
t_{\rm{cool,mix}}/t_{\rm{cc}} < 1 ,
\end{equation}
\end{ceqn}
where $t_{\rm{cool,mix}}$ is the cooling time of the intermediate temperature gas ($T\sim10^5$ K) formed by the mixing of the cool cloud with the hot corona and
\begin{ceqn}
\begin{equation}\label{eq:tcc}
t_{\rm{cc}} = \sqrt{\chi} r_{\rm{cl}}/v_{\rm{cl}}
\end{equation}
\end{ceqn}
is the cloud crushing time \citep{klein94}, where $\chi = \rho_{\rm{cool}} / \rho_{\rm{cor}}$. For our simulations, we found that $t_{\rm{cool,mix}}/t_{\rm{cc}}\sim 0.1$. This is larger than for simulations of clouds in the vicinity of the galactic disc (we evaluated for example that for the simulations of \cite{armillotta16} $t_{\rm{cool,mix}}/t_{\rm{cc}}\sim 0.03$) confirming that our clouds should be less prone to condensation, but is still lower than 1 and formally in the condensation regime. This is likely due to the simplicity of the criterion and the fact that, in contrast with the clouds studied by \cite{gronke18}, a cloud falling in the gravitational field of a galaxy halo cannot be entrained by the hot gas in the corona, as argued for instance by \cite{tan23}. In particular, \cite{tan23} have found that, in the presence of a gravitational field, the condition for cloud survival is more stringent than the condition by \cite{gronke18} and is given by:
\begin{ceqn}
\begin{equation}\label{eq:tgrow_4tcc}
t_{\rm{grow}}/4t_{\rm{cc}} < 1 ,
\end{equation}
\end{ceqn}
where $t_{\rm{grow}}$ is the time needed for a subsonic infalling cloud to grow (their equation 24) and the factor 4 was inferred empirically by \citet{tan23} from their simulations.
In Figure~\ref{fig:tgrowtcc} we show the comparison between $t_{\rm{grow}}$ and $4t_{\rm{cc}}$ in our simulation setup, as a function of the normalized distance from the central galaxy.\footnote{For this calculation, we used the cloud velocity as measured in the simulation, while cloud `radius' and density were derived for simplicity from the assumption of pressure equilibrium.} We can see that our clouds do not satisfy the criterion of equation~\eqref{eq:tgrow_4tcc} and therefore we do not expect the cool clouds to grow at any point in their evolution, in agreement with our findings. We should also stress here that the criteria expressed in equations \eqref{eq:tcool_tcc} and \eqref{eq:tgrow_4tcc} were derived for simulations that do not take into account the effects of thermal conduction, which is instead considered in our work.
 
The ability of cool IGM inflows to directly reach galaxies has long been addressed by large-scale cosmological simulations, finding a variety of results \citep[e.g.][]{dekel09,keres12,nelson16,wright20,hafen20}. In recent years, it has become clear that the  properties and fate of cool gas in the CGM are strongly dependent on numerical resolution. For instance, \citet{nelson20} studied the properties of the CGM of massive (${\rm M}_\star > 10^{11} \ {\rm M}_\odot$) galaxies at $z \sim 0.5$ in the cosmological simulation TNG-50 \citep{nelson19, pillepich19}. They found that gradually improving the baryonic mass resolution from ${\rm M}_{\rm bar} \sim 4.4 \times 10^7 \ {\rm M}_\odot$ down to ${\rm M}_{\rm bar} \sim 8.5 \times 10^4 \ {\rm M}_\odot$ leads to the gradual appearance of a large number of cool CGM clouds, improving the agreement with observational constraints on the covering factor of cool gas around massive galaxies \citep[e.g.][]{zahedy19}. Although the origin of cool CGM clouds in TNG-50 is not entirely clear, they appear to be long lived and the majority of them fall all the way down to the central galaxy, in contrast with our findings. The discrepancy may be partially due to the different redshift ($z \sim 0.5$ vs our $z = 0$) and higher halo masses (typical ${\rm M}_{\rm h} > 10^{13} \ {\rm M}_\odot$ vs our maximum ${\rm M}_{\rm h,max} = 10^{12.3} \ {\rm M}_\odot$), but it is possible that an important role is also played by the still limited resolution in TNG-50. In fact, even the best mass resolution achieved by TNG-50 (${\rm M}_{\rm bar} \sim 8.5 \times 10^4 \ {\rm M}_\odot$) is four orders of magnitude worse than in our idealized simulations (about $4 \ {\rm M}_\odot$, calculated for cool gas at $z= z_{\rm ev}$).

The widely recognized importance of resolution for CGM studies has recently motivated the development of a family of zoom-in cosmological simulations specifically designed to achieve an improved uniform spatial resolution in the CGM, with cell sizes as small as 0.5-1.0 kpc \citep{vandevoort19, peeples19}. This is better than the TNG-50 resolution (for relevant values of the gas densities and in particular in the halo outskirts), but still worse than in our simulations. Applying this methodology to a MW-like simulated galaxy, \citet{zheng20} found that infalling cool clouds disappear in the corona before reaching the central disc, typically at a distance of $z \sim 20$ kpc from the centre. This is in better agreement with our findings, although we find that evaporation happens at significantly larger distances (see Table \ref{tab:resultsEvap}). Once again, the discrepancy could be attributed to the still limited spatial resolution of \cite{zheng20} simulations. This explanation is supported by our own resolution study in Section \ref{convergence}, where we have found that a spatial resolution worse than about 62.5 pc is already sufficient to artificially increase the survival time of cool clouds. In conclusion, we argue that current cosmological simulations, including zoom-in simulations with forced uniform spatial CGM refinement, have not yet reached the resolution needed to make reliable predictions on the fate of accreting cool clouds.

\subsection{What governs the evolution of infalling cool clouds?}\label{subsec:processes}
Although a full first-principle understanding of the processes acting in our simulations is beyond the scope of this work, we propose here a description of the main physical effects that plausibly play important roles in the evolution of our simulated infalling cool clouds. In particular, a close inspection of some of our simulations (MWFID and MW100, see Table \ref{tab:inCond}) suggests that a sequence of three crucial events determines the evolution and eventual disruption of our cool clouds.

First, an infalling cool cloud loses its spherical shape and is flattened due to the ram pressure exerted by the hot gas. This process seems to occur on approximately one cloud crushing time (equation \ref{eq:tcc}). For simulation MW100, in which the velocity of the cool gas is roughly constant with time ($v(t) \sim v_{\rm in} = 100 \; {\rm km} \ {\rm s}^{-1}$, see Figure \ref{fig:profilesvel}), the crushing time is $t_{\rm cc} = 300 \ {\rm Myr}$ , which corresponds well to the time after which the cloud is fairly flattened in this simulation. For MWFID, the comparison is less straightforward, as the cloud velocity (and therefore formally $t_{\rm cc}$) is not constant but increases (decreases) with time (see Figures \ref{fig:profilesvel} and \ref{fig:tgrowtcc}). In this case, we introduce a critical time, defined by the condition $t_{\rm crit} = t_{\rm cc}(t_{\rm crit})$, before which ram pressure effects should be negligible and after which flattening should be completed in roughly one further crushing time. In MWFID, we calculate $t_{\rm crit} \sim 0.65 \ {\rm Gyr}$. In agreement with our simple expectation, the cloud remains roughly spherical for the first $\sim$ 0.65 Gyr, after which it starts showing significant deformations, and is fully flattened after further $\sim$ 0.65 Gyr, i.e.\ $\sim 2 t_{\rm crit} = 1.3 \ {\rm Gyr}$ from the start of the simulation.

Second, the flattened cloud is fragmented along the horizontal direction (i.e. perpendicular to the cloud's motion) into a few major pieces. We attribute this mainly to the Rayleigh-Taylor instability, which is expected to lead to fragmentation in the horizontal direction on a typical time-scale $t_{\rm RT} = \sqrt{r_{\rm cl}/\pi g_{\rm eq}}$, where $g_{\rm eq}$ is the equivalent gravitational field felt by the cloud in its own non-inertial rest-frame and is readily shown to be equal to the acceleration contributed by the ram pressure (a.k.a.\ the `drag force'). For MWFID, we find $t_{\rm RT} \sim 200 \ {\rm Myr}$ (evaluated after flattening), similar to the observed time required for fragmentation in this simulation (see also Figure \ref{fig:snapshots}). In addition to the Rayleigh-Taylor instability, the Kelvin-Helmholtz instability is also expected to contribute to fragmentation (on a typical timescale $t_{\rm KH} \sim t_{\rm cc}/2\pi$), in particular at the edges of the cloud, where shear motions are important.

Third, the cloud fragments produced by hydrodynamical instabilities are further compressed and sometimes further fragmented by the combined action of the pressure of the hot gas (which increases with decreasing $z$ due to the vertical stratification of the corona) and turbulence, which develops as a consequence of instabilities. Turbulence also creates transient cool gas structures on a cascade of spatial scales. As long as fragments, or transient structures, appear at progressively smaller scales, the conduction time decreases in quadratic proportion (see equation \ref{eq:tcond}) and thence the cool gas structures become increasingly prone to evaporation through thermal conduction. In our simulations, we observe structures as small as 200 pc being evaporated on a time-scale of 50 Myr or less, in good agreement with equation \eqref{eq:tcond} for these spatial scales and the cool gas density at $z \sim z_{\rm ev}$. Note that cool gas structures do not need to be small in all directions to be efficiently evaporated. In particular, we observe that some of the cloud fragments are highly anisotropic (see also Figure \ref{fig:SnapshotResolutions}), with filamentary structures that are as long as 5 kpc but as thin as 200 pc and, accordingly, are destroyed by thermal conduction in a few tens of Myr.

In conclusion, we confirm that thermal conduction is the key process responsible for the final evaporation of the cool gas. However, it does not determine the typical timescale for the overall survival of cool gas clouds, which is rather set by other hydrodynamical effects, such as ram pressure, the Rayleigh-Taylor instability and the Kelvin-Helmholtz instability. Thermal conduction, however, determines the typical size that cool gas structures must reach, through fragmentation or turbulent cascades, before they can be efficiently evaporated. This is also in agreement with Section \ref{convergence}, where we have found that sufficiently resolving the Field length (i.e.\ the fundamental spatial scale for thermal conduction), is required to reach numerical convergence.

\subsection{Implications for gas feeding of star forming galaxies}\label{implications}
Cosmological models predict that baryonic matter is continuously accreting into the halos of galaxies. Using the prescription of \cite{correa15a,correa15b}, which we also used in Section~\ref{coolCloudsProps} to define the initial conditions of our cool infalling clouds, we estimated the total accretion of gas into the halo of a M31- and a MW-like galaxy to be respectively 11.5 and 8 $M_{\odot}\ \rm{yr}^{-1}$. However, the star formation rates (SFRs) observed in M31 and the MW are respectively less than 1 $M_{\odot}\ \rm{yr}^{-1}$ \citep[e. g.][]{rahmani16} and a few $M_{\odot}\ \rm{yr}^{-1}$ \citep[e.g.][]{diehl06, blandhawthorn16}. It is therefore expected that the majority of this accreting material either will not reach the central galaxy, or its accretion must be continuously compensated by massive outflows. Our work supports the first option, as we have found that thermal conduction suppresses the direct accretion of IGM clouds onto disc galaxies with stellar masses ranging from $10^{10.1}$ to $10^{10.7} M_{\odot}$, at least at low redshift, and therefore this gas does not survive to reach the central galaxy.

However, this scenario raises, at least at first sight, the opposite problem: if no accretion is taking place, how can we explain the observed levels of star formation? It has long been known \citep[e.g.][]{kennicutt83} that, in the absence of accretion of cold gas from the surrounding environment, the interstellar medium of star-forming galaxies would be depleted in about one or at most a few Gyr \citep[e.g.][]{fraternali12, saintonge13} and would therefore not be able to sustain the observed levels of star formation. Moreover, chemical evolution models \citep[e.g.][]{schonrich09} specifically require the continuous accretion of low-metallicity gas, disfavouring feeding by internal mechanisms such as stellar mass loss \citep[e.g.][]{leitner11}.

One natural explanation would be that the accretion happens through the cooling of the hot corona \citep[e.g.][]{white91,fukugita06}. Recently, \cite{stern19} proposed that in galaxies with halos similar to that of the Milky Way the corona might be virialized in the outer halo and then cool efficiently at small radii, thence providing cold gas to the galaxy. This is also seen in the FIRE-2 simulations \citep[e.g.][]{hafen22}, where this type of accretion seems to correlate with the formation of the galaxy thin disc. 

The exact physical processes that trigger the cooling of the corona at the interface with the disc are still a matter of debate. In this context, one mechanism that could be able to sustain the galaxy star formation is the accretion driven by the galactic fountain \citep{shapiro76,bregman80,fraternali06, fraternali08}. In this scenario, cold gas expelled from the disc by supernova explosions drives the condensation of the hot corona \citep[through the mechanisms mentioned in Section \ref{disc:comparison}, e.g.][]{marinacci10accr,armillotta16}, eventually falling back to the disc and thence providing further fuel for star formation. This model reproduces various observational constraints of cold extraplanar gas and intermediate- and high-velocity-clouds in the Milky Way \citep[see][]{marasco12,fraternali15,marasco22}, with a net accretion rate consistent with what is needed to sustain observed levels of star formation \citep[see][and references therein]{fraternali17} and possibly also the angular momentum needed to sustain the observed rate of inside-out growth \citep{li23}. This `positive feedback' scenario for gas accretion may therefore explain the continuous star formation of disc galaxies, without the need of direct accretion of cool gas from the IGM, in line with the results of this paper.

Finally, other processes have also been proposed that could lead to the formation (and the subsequent accretion) of cool gas in the inner parts of the halo, including in particular the condensation of clumps of cool gas out of the corona due to thermal instabilities \citep[e.g.][]{sharma12,voit17,sormani19,nelson20} and the stripping of cool gas from satellites \citep[e.g.][]{grcevich09, marasco16, alcazar17, johnson18, hafen19}. If the cool gas originates close enough to the central galaxy, these clouds may survive their infall and feed the central star formation. To accurately estimate the survival time of such clouds one would need tailored simulations, whose implementation is outside the scope of this work.

\subsection{Implications for the modelling of cool CGM absorbers}\label{implications:absorbers}
  \begin{figure}
   \includegraphics[clip, trim={0cm 0cm 0cm 0cm}, width=\linewidth]{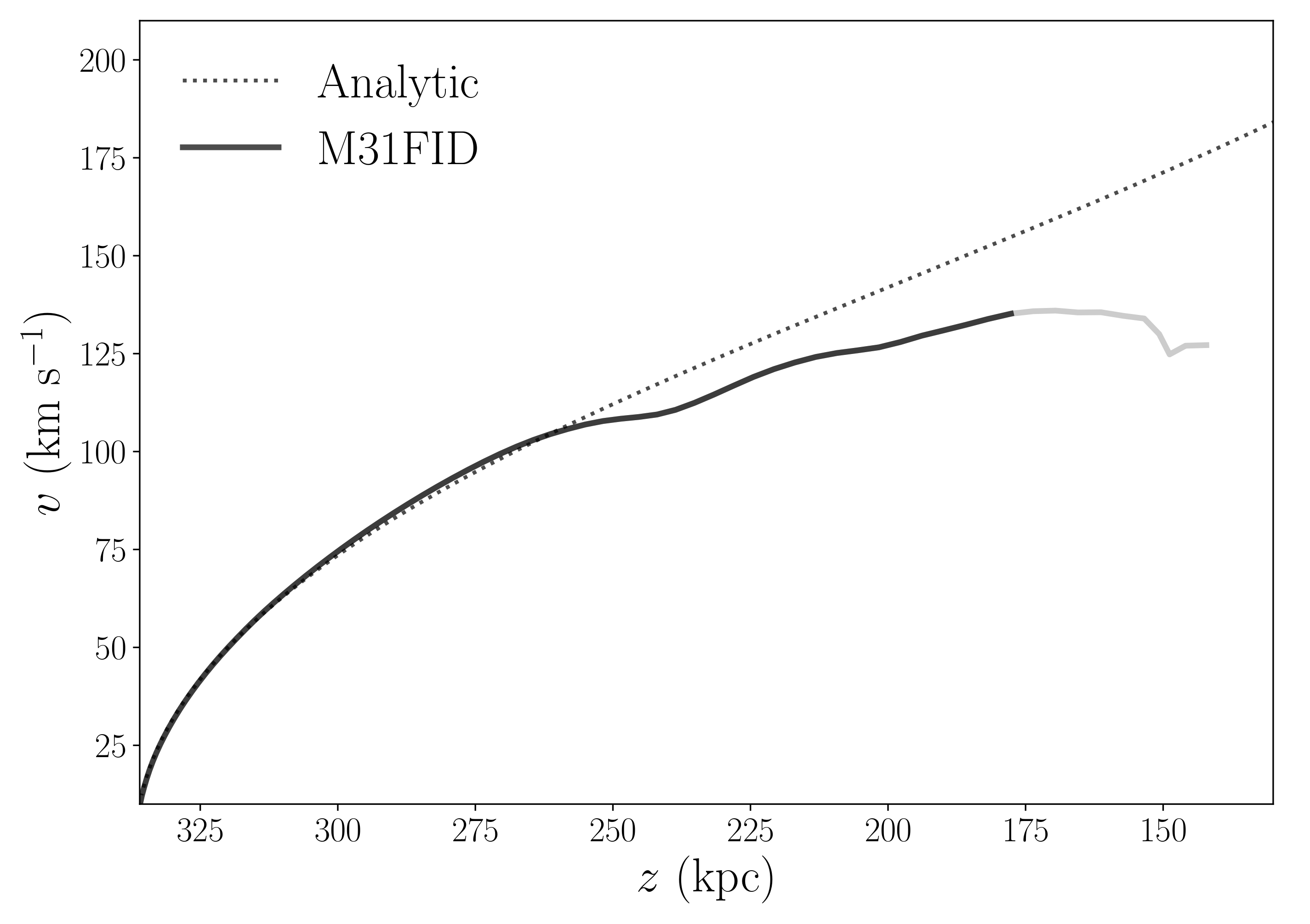}
   \caption{Simulated and analytic velocity profiles as a function of the galactocentric distance for a cool cloud infalling through the halo of the Andromeda galaxy. Dotted curve: analytical prediction that takes into account the gravitational force and the drag force exerted by the hot corona \citep[see][]{afruni22}; solid curve: mass weighted cool gas velocity in the simulation M31FID, with a higher level of transparency for $z<z_{\rm{ev}}$. The analytical model traces the simulation perfectly until $z\sim270$ kpc and then systematically overpredicts it by up to 25 km s$^{-1}$.}
              \label{fig:vansims}%
    \end{figure}
In our previous works \citep[][]{afruni19,afruni20,afruni22}, we have used semi-analytical models to infer the dynamics and origin of the observed cool circumgalactic clouds around low-redshift galaxies, by comparing model predictions with observational data of the cool CGM \citep[see][]{werk13,borthakur15,zahedy19,lehner20p}. With our new hydrodynamical simulations we were able to study in a much greater detail the survival of the clouds and in general their interaction with the hot corona, which, in the semi-analytic approach, could be treated only in an approximated way. 
It is therefore important to explore whether and how our current findings  affect the results of our previous semi-analytical models. It is particularly interesting to compare the results that we obtained here with the work of \cite{afruni22}, where we have modelled in detail the cool CGM around M31 and found that it is consistent with accretion of metal-poor clouds originated from the IGM. 

In order to keep the number of free parameters to a minimum, \cite{afruni22} made a number of simplifying assumption, including (i) that the clouds are spherical at all times (ii) that they do not lose mass during their journey and (iii) that the velocity of the cool gas evolves following a simple equation of motion, including the effects of gravity and a simple parametrization of the drag force due to the interaction with the corona. The first two assumptions are evidently violated by the results of our simulations. Nonetheless, we have found that their impact on the kinematics (the key observable on which the inference in \citealt{afruni22} was based), is only minor. To illustrate this, we show in Figure~\ref{fig:vansims} a comparison between the velocity predicted by \cite{afruni22} (where the motion of the cloud is governed by gravity and the drag force) and the cool gas velocity found in the simulation M31FID (see Table~\ref{tab:inCond} and Section~\ref{resVelocity}), both as a function of the distance from the central galaxy. The profiles agree perfectly well for $z>270$ kpc. For distances between 270 kpc and 150 kpc the simulated velocity profile is lower than the analytic prediction, but differences are mostly smaller than 25 km s$^{-1}$. The reasons for these (moderate) discrepancies at late times are likely related to the assumptions listed above and in particular (i) the simulated cloud loses mass with time, leading to a progressively higher deceleration for a given drag force exerted by the hot corona and (ii) because of the flattening (due to the ram pressure), the simulated cloud tends to have an increased cross section in the direction of the infall (see Figure~\ref{fig:SnapshotResolutions}), which in turn increases the drag force (and thence the deceleration) compared to the analytical model.
 
We will show in a separate work (Afruni et al., in preparation) that, taking into account the (slightly) different velocity and the (significant) mass loss in the inner regions, the model of \cite{afruni22} still provides an acceptable description of the observational data of \cite{lehner20p}. The basic reason is that (as already pointed out in \citealt{afruni22}) the semi-analytical model also predicted that the majority of the observed clouds are in fact located at large intrinsic (i.e. deprojected) distances from the disc of M31, where the impact of detailed hydrodynamical effects is small. The possible implication is that, although the cool CGM of M31 may mostly originate from IGM accretion, this accretion may not be responsible for the ongoing star formation in this galaxy, in agreement with what discussed in Section \ref{implications}.

In perspective, we believe that the combined use of detailed hydrodynamical simulations and flexible semi-analytical models \citep[see also][]{fielding22} is a promising path to understand the observed CGM, as well as its connection with the evolution of galaxies.

\section{Summary and conclusions}\label{conclusions}
In this work, we used 3D high-resolution hydrodynamical simulations to study the survival of clouds accreted from the IGM as part of the cosmological accretion of gas into the halos of disc galaxies at low redshift. Our main goal was to explore whether these clouds are able to reach the central galaxies and constitute a viable source to feed their star formation. We included in our simulations the effects of gravity, thermal conduction, radiative cooling and an ionizing UV background. We ran all our experiments using AMR, reaching a maximum resolution in our fiducial simulations of 62.5 pc (the initial clouds' diameter is about 160 times the cell length), which allows us to resolve the Field length and reach numerical convergence. We explored three different environments, representing the halos of M31- , MW- and Sub-MW- like galaxies, with virial masses equal to respectively $10^{12.3}, 10^{12.1}$ and $10^{11.9}\ M_{\odot}$. For each of these halos we explored different initial conditions, varying the initial cloud mass and velocity, the density of the corona and the value of the suppression factor of thermal conduction. Our main findings are the following:
\begin{enumerate}
    \item all our simulations show that the cool IGM clouds lose the vast majority of their mass, evaporating in the surrounding hot corona, after an evaporation time that varies from 0.75 to 2 Gyr, at distances from the central galaxy that vary from 0.52 to 0.78$r_{\rm{vir}}$. These clouds are therefore unable to reach the disc and directly contribute to its star formation, but rather contribute to feed the corona;
    \item we do not find any particular dependence in the behavior of the cool cloud with the choice of the halo mass and the suppression factor of the thermal conduction; 
    \item clouds embedded in more massive (denser) coronae or with higher initial infall velocities tend to evaporate faster. This effect is likely related to the fact that these clouds need to be less massive than in our fiducial setup to satisfy observational constraints on the covering fraction of the cool CGM;
    \item resolving the Field length with 5-7 cells is sufficient to reach numerical convergence in the evolution of the cool gas mass. Conversely, cosmological and zoom-in simulations currently achieve mass (spatial) resolutions that are typically at least 3 orders (1 order) of magnitude worse than in this study, leading to an artificial increase in the survival time of cool gas inflows against evaporation.
\end{enumerate}

The results of our simulations indicate that cold accretion from the IGM is not a viable way to feed the star formation of disc galaxies at low redshift, as the fate of the cool infalling clouds is to evaporate into and feed the hot corona. This favours a scenario in which accretion of cold gas onto present-day massive star-forming galaxies proceeds from condensation of the corona (`hot mode') in its very inner regions, where it is in contact with the galactic disc. Such process can be either spontaneous or mediated by a galactic fountain.
\section*{Acknowledgements}
The authors would like to thank Lucia Armillotta, for providing the CLOUDY tables that have been used to calculate the gas cooling and heating rates. We are grateful to the anonymous referee for insightful and interesting comments. AA acknowledges the financial support of the Joint Committee ESO-Chile grant. GP acknowledges support from the Netherlands Research School for Astronomy (Nederlandse Onderzoekschool Voor Astronomie, NOVA). The analysis of the simulation outputs presented in this paper has been done using the \textit{yt} python toolkit \citep[][]{yt11}.
\section*{Data Availability}
The simulation data presented in this article are available upon reasonable request to the corresponding author.




\bibliographystyle{mnras}
\bibliography{biblio} 







\bsp	
\label{lastpage}
\end{document}